\documentclass[prd,twocolumn,aps,floatfix,onecolumn,12pt]{revtex4}

\usepackage{amsmath}
\usepackage{graphicx}
\usepackage{dcolumn}
\usepackage{bm,bbm}

\def\be{\begin{equation}}
\def\ee{\end{equation}}    
\def\baray{\begin{eqnarray}}
\def\earay{\end{eqnarray}}

\def\2pi{\left(2\pi\right)}
\def\ve{\overrightarrow}
\def\hf{\frac{1}{2}}

\def\dxf{\frac{\partial f}{\partial\phi}}
\def\dyf{\frac{\partial f}{\partial\overline{\phi}}}
\def\dxdyf{\frac{\partial^2 f}{\partial\phi\partial\overline{\phi}}}
\def\dxg{\frac{\partial g}{\partial\psi}}
\def\dyg{\frac{\partial g}{\partial\overline{\psi}}}
\def\dxdyg{\frac{\partial^2 g}{\partial\psi\partial\overline{\psi}}}

\begin{document}

\title{Small cosmological signatures from multi-brane models}
\author{Carlo R.~Contaldi}
\email{c.contaldi@imperial.ac.uk}
\author{Gavin Nicholson}
\email{gavin.nicholson05@imperial.ac.uk}
\author{Horace Stoica}
\email{f.stoica@imperial.ac.uk}
  \affiliation{Blackett Laboratory, Imperial College London, SW7 2AZ, UK}
\date{\today}

\begin{abstract}We analyse the signatures of brane inflation models with
  moduli stabilisation. These are hybrid inflation models with a
  non-trivial field-space metric which can induce complex trajectories
  for the fields during inflation. This in turn could lead to
  observable features on the power spectrum of the CMB fluctuations
  through departures from near scale invariance or the presence of
  isocurvature modes. We look specifically at multi-brane models in
  which the volume modulus also evolves. We find that the signatures
  are highly sensitive to the actual trajectories in field space, but
  their amplitudes are too small to be observable even for future high
  precision CMB experiments.
\end{abstract}

\maketitle

\section{Introduction}

The early universe gives us the best possibility to test string theory
models. Brane-anti-brane \cite{Dvali:1998pa,Burgess:2001fx,GarciaBellido:2001ky,Jones:2002cv}
and modular inflation models \cite{BlancoPillado:2004ns,Conlon:2005jm}
give predictions that are largely compatible with the present cosmological
data. Therefore, in order to differentiate between the various models we
have to look at their predictions beyond the usual set 
of experimentally determined parameters (the amplitude of the density 
perturbations, the spectral index and the running of the spectral
index). 

With the advent of high precision cosmological Cosmic Microwave
Background (CMB) and Large Scale Structure (LSS) data the possibility
of characterising the initial perturbation spectrum beyond these
simplest spectral parameters is now a reality. Efforts to reconstruct
the initial perturbation spectrum with model-independent 'inversion'
techniques are already under way (see e.g. \cite{reconstruct}) and
have yielded some tantalizing hints of structure in the
spectrum. There is a lack of power on the largest scales
\cite{quadrupole} and some indications of oscillations on intermediate
scale \cite{oscillations}. Arguably, the statistical significance of
these effects are not sufficiently strong to motivate a departure from
the simplest phenomenological models of single-field, slow roll
inflation but an accurate analysis of observable features in
theoretically driven scenarios is certainly warranted.

A generic feature of models of inflation within warped
compactification mechanisms is the presence of many extra degrees of
freedom in addition to the field driving inflation. Depending on the
masses and couplings of the extra fields their presence can induce
non-trivial trajectories in the field space. This can result in
observable effects such as radically broken scale invariance, presence
of isocurvature (or entropy) perturbations and enhanced
non-Gaussianity of the perturbations close to points of high
acceleration in the trajectories \cite{ng}.

A non-vanishing isocurvature contribution to the total initial
perturbations is a particularly interesting possibility since tight
limits on any such contribution will be available when CMB
polarization measurements becomes precise. The polarization data will
also help break fundamental degeneracies in our ability to accurately
constrain any broken scale invariance beyond a simple running of the
spectral index.

In this work we specifically look at multi-brane models
\cite{Cline:2005ty}. These models are very promising since they
naturally avoid a number of significant fine-tuning problems present in the
simplest brane inflation models. Firstly they
naturally produce a larger number of $e$-folds via the assisted
inflation mechanism \cite{Liddle:1998jc}, and secondly, they offer the
possibility that the inflaton potential is flattened
dynamically. 

In these models, all branes are initially initially in a local minimum
and as branes tunnel through the potential barrier and annihilate with
anti-branes the barrier is lowered dynamically until it completely
disappears. The potential becomes monotonic and very close to flat
once a sufficient number of branes-anti-brane pairs have annihilated
and the remaining branes will roll down the potential before
annihilating. Quantum fluctuations will cause the rolling branes to
start from slightly different initial positions and therefore follow
different trajectories. We study the signatures that these models give
which may be detected through cosmological observations.

All models under consideration feature moduli stabilisation for both
the shape \cite{Giddings:2001yu} and volume \cite{Kachru:2003aw}
moduli and anti-branes needed to lift the anti-deSitter vacuum
generated by the stabilising mechanism to a deSitter one during
inflation. The bulk contains a warped throat, the anti-branes being
located at the bottom of it, and the mobile branes roll down this
throat while inflation takes place. 

Specifically, we follow the evolution of a number of perturbation
modes for a set of scalar fields (in this case the positions of the
mobile branes) coupled to gravity both inside and outside the Hubble 
horizon. The non-trivial shape of the potential and K\"ahler metric
leads to a residual evolution of the scalar fields after horizon
crossing. We evolve the perturbations of the scalars coupled to
gravity using the Mukhanov-Sasaki \cite{Mukhanov:1990me,Sasaki} variables and decompose the
perturbations into adiabatic and entropy ones. We finally compute the
separate spectra for the adiabatic and entropy perturbations and we
find that while the adiabatic perturbations are insensitive to the
trajectories followed by the many inflaton fields, the entropy ones
are highly dependent on the trajectory. However, their amplitude is 
much too small to lead to observable features in the measured CMB
spectrum. 

The study performed here applies to a larger class of models, namely
the ``inflection point'' models \cite{Baumann:2007np}, where the mass
of the inflaton(s) is large, except for a very small region  around
an inflection point of the potential seen as a function of one
inflaton field at a time (keeping all other fields constant). Most of
the $e$-folds are coming from this small region of the field space where
the inflaton trajectory can be very well approximated by a straight 
line therefore leading to a very small amplitude for the density 
perturbations.

\section{The model}

As mentioned in the introduction, the model we study here covers
a more general class of brane-anti-brane inflation models.
The important features are:
\begin{itemize}
\item Stabilisation of the complex structure moduli via fluxes 
  \cite{Giddings:2001yu} and of the volume modulus via non-perturbative 
  effects \cite{Kachru:2003aw,Kallosh:2004yh})
\item The presence of anti-branes that lift the vacuum to deSitter
  during inflation.
\item A number of mobile branes that will roll down the warped throat
  and annihilate with the anti-branes, ending inflation. 
\end{itemize}
In these scenarios the dynamics of the inflaton is determined by the 
brane anti-brane interaction and by the mechanism responsible for the 
stabilization of the volume modulus. The vacuum energy responsible for 
inflation is provided by the (warped) anti-brane tension. 
\begin{figure}[thbp]
\begin{center}
\includegraphics[width=8cm,angle=0]{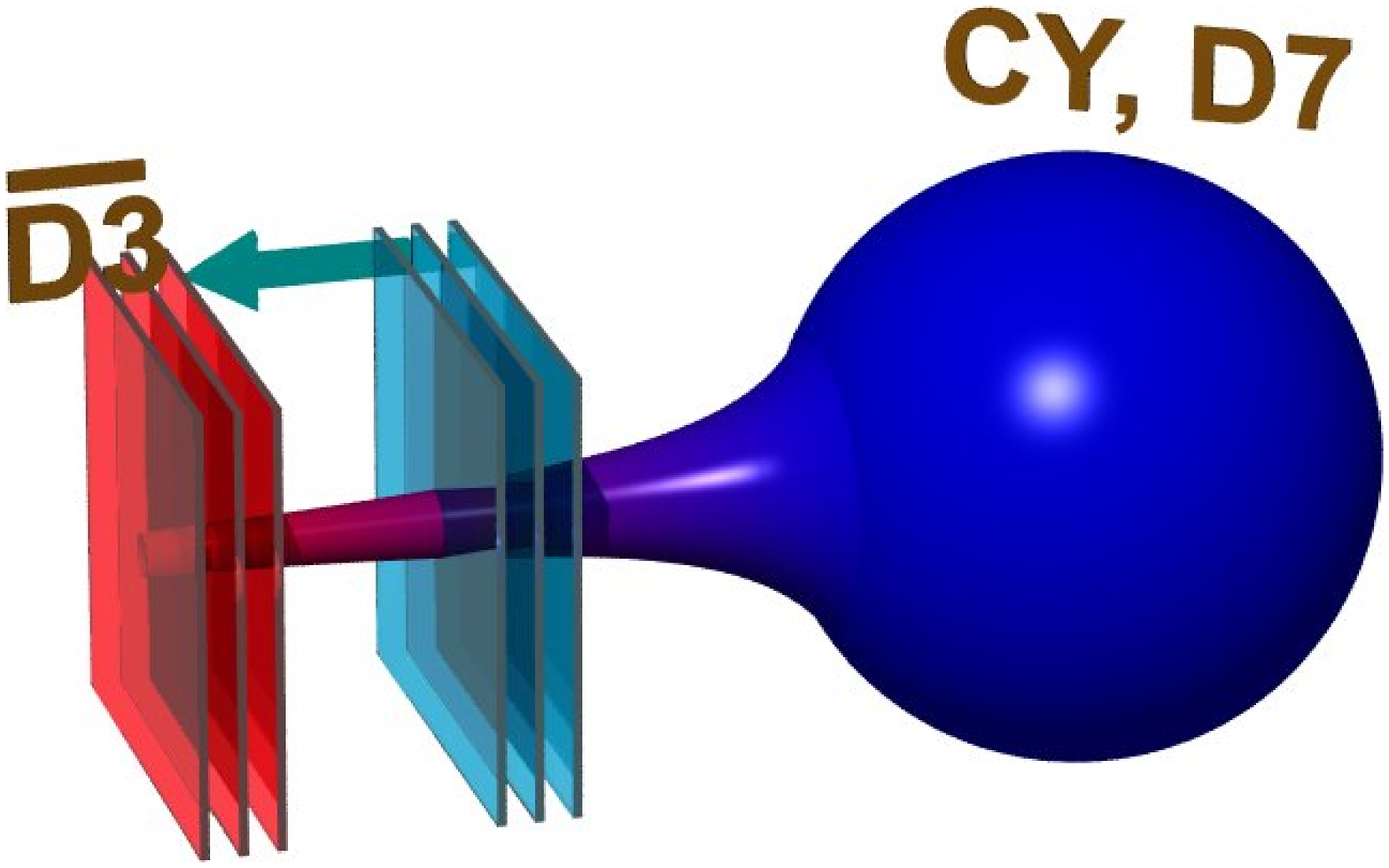}
\hspace{6pt}
\includegraphics[width=8cm,angle=0]{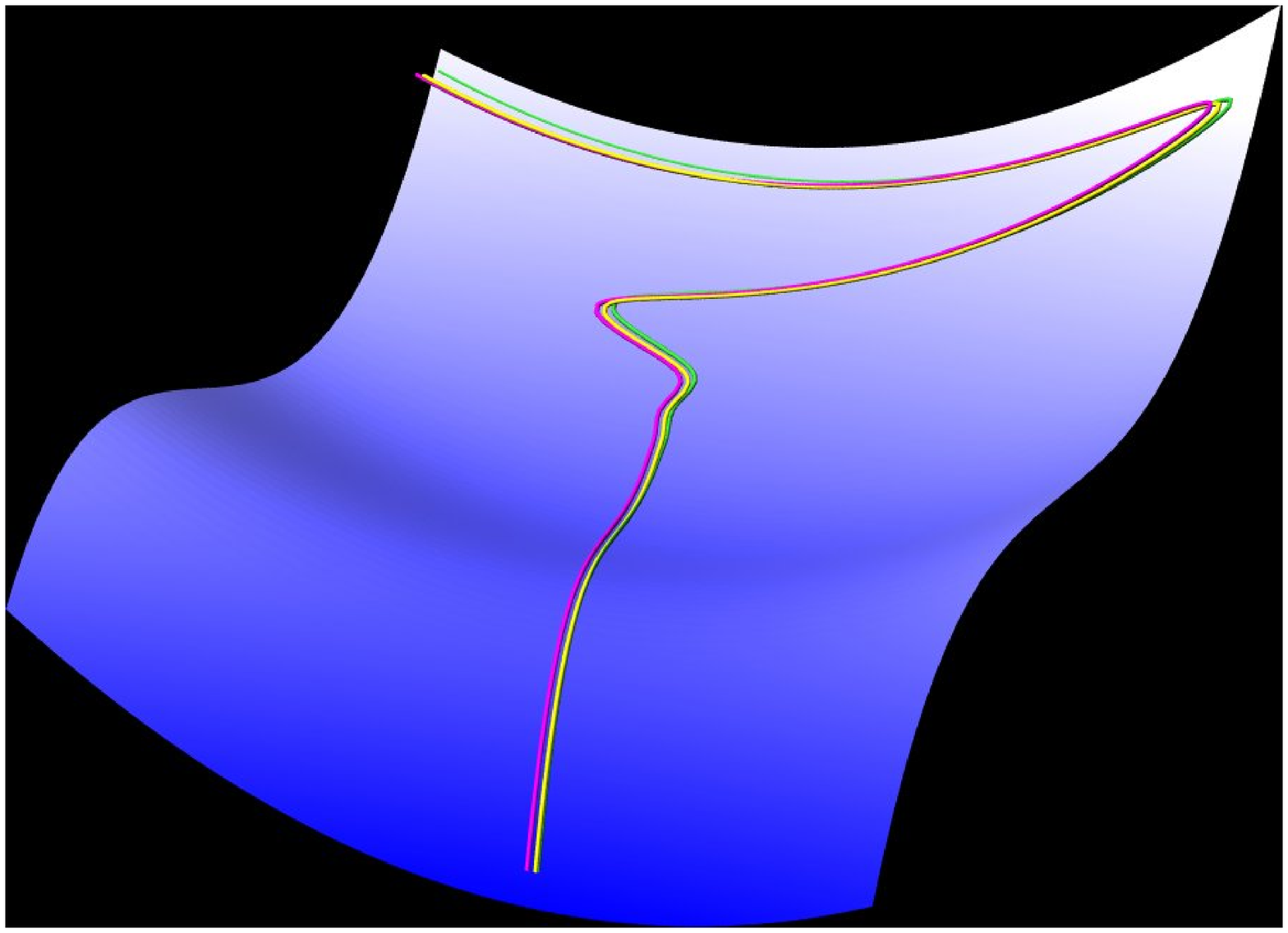}
\caption{Brane setup and field trajectories. The potential features 
a local minimum for the volume modulus and an inflection point for 
the inflaton, the brane-anti-brane separation. 
\label{trough}}
\end{center}
\end{figure}
The stabilization of the volume modulus is done by $D7$ branes spanning 
the 3 large space dimensions and wrapping four-cycles of the compact space. 
Gaugino condensation in the stacks of wrapped $D7$ branes generates a 
non-trivial superpotential for the volume modulus, which in the simplest case 
takes the form given in \cite{Kachru:2003aw}. 
\be
W = w_{0}+Ae^{-iaT}
\label{KKLT_pot}
\ee
The non-trivial embedding of the $D7$ branes in the compact 
space makes the coefficient $A$ dependent on the position of the mobile 
$D3$ brane \cite{Baumann:2006th}. An additional dependence of the scalar 
potential derived from Eq.(\ref{KKLT_pot}) on the position of the mobile 
brane comes from the non-trivial K\"ahler metric \ref{Kahler_Metric}.  
The resulting effects of the moduli stabilization and the uplifting of the 
vacuum from anti-deSitter to deSitter by adding anti-$D3$ branes give a 
potential that is generally too steep for inflation, the value of the
slow-roll parameter $\eta$ being $\eta = 2/3$ \cite{Kachru:2003sx}.
However, when taking into account the competing effects of the 
moduli stabilization and anti-brane attraction, the resulting  potential 
for the mobile brane features an inflection point 
\cite{Burgess:2004kv,Cline:2005ty,Baumann:2007np}. The vanishing of the
$\eta$ parameter at this point makes it possible for inflation to work, 
most of the $e$-folds coming from a small region of field space. 

Apriori, such models require a large amount of fine-tuning, as the slope 
of the potential, and therefore the $\epsilon$ slow-roll parameter, at the 
inflection point must be very small. However, multi-brane models 
offer the possibility to alleviate this fine-tuning via the dynamical 
flattening mechanism for the inflaton potential. In what follows we will
study the cosmological signatures of multi-brane models going beyond the 
calculations of amplitude of the density perturbations, the spectral index and
it running to determine what unique signatures one can expect to be 
detectable in cosmological observations.

\subsection{The K\"ahler potential}
The K\"ahler potential proposed by DeWolfe and Giddings
\cite{DeWolfe:2002nn} has the following form:
\begin{equation}
{\mathcal K}=-3\log\left(r\right)=
-3\log\left(T+\overline{T}-f\left(\varphi^i,\overline{\varphi}^i\right)\right),
\label{K_pot}
\end{equation}
where we will assume that for a multi-brane setup the function $f$ 
decomposes on a sum of functions, each one depending on one brane position $\varphi^i$:
\begin{equation}
\label{KPot}
{\mathcal K} = -3\log\left(T+\overline{T}-
\sum_{i}f_i\left(\varphi^i,\overline{\varphi}^i\right)\right).
\end{equation}
The calculations in this paper use the simplest form of the functions $f_i$:
\begin{equation}
f_i\left(\varphi^i,\overline{\varphi}^i\right) = \varphi^i\overline{\varphi}^i.
\end{equation} 
One can justify this assumption by taking into account the fact that
most $e$-folds of inflation are produced around the inflection point of
the potential and therefore the functions $f_i$ can be expanded to the 
lowest order in the brane positions. 

In the general case the K\"ahler metric derived  from the K\"ahler 
potential of Eq.(\ref{KPot}) has the form:
\begin{equation}
\label{Kahler_Metric}
{\mathcal K}_{a\overline{b}} = 
\frac{3}{r^2}\left(\begin{array}{ccccc}
1 & -\frac{\partial f_1}{\partial\varphi_1} & -\frac{\partial f_2}{\partial\varphi_2} & 
\cdots & -\frac{\partial f_n}{\partial\varphi_n} \\
-\frac{\partial f_1}{\partial\varphi_1^*} & 
\frac{\partial f_1}{\partial\varphi_1}\frac{\partial f_1}{\partial\varphi_1^*}+ 
r\frac{\partial^2 f_1}{\partial\varphi_1\partial\varphi_1^*} &
\frac{\partial f_1}{\partial\varphi_1^*}\frac{\partial f_2}{\partial\varphi_2} & 
\cdots & \frac{\partial f_1}{\partial\varphi_1^*}\frac{\partial f_n}{\partial\varphi_n} \\
\vdots & \vdots & \vdots & \ddots & \vdots \\
-\frac{\partial f_{n-1}}{\partial\varphi_{n-1}^*} 
&\frac{\partial f_{n-1}}{\partial\varphi_{n-1}^*}\frac{\partial f_1}{\partial\varphi_1}  
&\frac{\partial f_{n-1}}{\partial\varphi_{n-1}^*}\frac{\partial f_2}{\partial\varphi_2} 
&\cdots 
&\frac{\partial f_{n-1}}{\partial\varphi_{n-1}^*}\frac{\partial f_n}{\partial\varphi_n} \\
-\frac{\partial f_{n}}{\partial\varphi_{n}^*} 
&\frac{\partial f_n}{\partial\varphi_n^*}\frac{\partial f_1}{\partial\varphi_1}
&\frac{\partial f_n}{\partial\varphi_n^*}\frac{\partial f_2}{\partial\varphi_2}
&\cdots 
&\frac{\partial f_n}{\partial\varphi_n^*}\frac{\partial f_n}{\partial\varphi_n}+
r\frac{\partial^2 f_n}{\partial\varphi_n\partial\varphi_n^*}
\end{array}\right)
\end{equation}
The inverse can also be calculated in the general case (see appendix \ref{F_term_details}).
We want to write the kinetic energy in terms of the real and imaginary parts of the 
fields \cite{Burgess:2004kv}. The imaginary parts will play no role in our subsequent 
analysis as they will roll very quickly to their respective minima and these minima do 
not shift during the rolling of the real parts \cite{Kallosh:2004yh}.
We write the volume modulus as $T = \sigma + {\mathbbm i}\omega$
and the brane positions as $\varphi^{i} = \phi^{i}+{\mathbbm i}\chi^{i}$
which allows us to read off the field space metric and it inverse for
the $\sigma$ and $\phi^{i}$ fields:
\begin{equation}
G_{ij}=\frac{6}{r^2}\left(\begin{array}{cc}
1 & -\phi^i \\
-\phi^i & r\delta_{ij} + \phi^i\phi^j 
\end{array}\right)
\;\; ,
\;\;\;\;
G^{ij}=\frac{r}{6}\left(\begin{array}{cc}
2\sigma & \phi^{i} \\
\phi^{i} & \delta_{ij} 
\end{array}\right)
\end{equation}

\subsection{The potential}
In this section we describe the various contributions to the scalar potential. They 
are the F-term responsible for the stabilization of the volume modulus, the anti-brane
contribution that uplifts the resulting anti-deSitter minimum to a deSitter one, and 
the brane-anti-brane attractive potential which gives a non-trivial evolution to the
inflatons. The F-term is calculated with the usual formula:
\begin{equation}
V_F=e^{\mathcal K}\left(K^{\overline{a}b}\overline{D_{a}W}D_{b}W-3\left|W\right|^2\right),
\end{equation}
where ${\mathcal K}$ is given in Eq. (\ref{KPot}).
Let us take for example a scenario with two mobile branes; the K\"ahler potential is:
\begin{equation}
K=-3\log\left(r\right)=
-3\log\left(T+\overline{T}
-f_{1}\left(\phi^{1},\overline{\phi^{1}}\right)
-f_{2}\left(\phi^{2},\overline{\phi^{2}}\right)\right).
\end{equation}
We take the superpotential to be the racetrack one, this way the
vacuum with no anti-brane present is already Minkowski \cite{Kallosh:2004yh}:
\begin{equation}
\label{racetrack_W}
W=w_0+Ae^{-aT}+Be^{-bT}.
\end{equation}
If we take the minimal form for the functions 
$f_{1}\left(\phi^{1},\overline{\phi^{1}}\right)$ and 
$f_{2}\left(\phi^{2},\overline{\phi^{2}}\right)$ and the general expression for the 
F-term, Eq. (\ref{general_F_term}), the resulting potential has a very simple form:
\baray
V_F&=&\frac{e^{-a\sigma}}{3r^2}\left[aA^2e^{-a\sigma}\left(3+a\sigma\right)+
aAw_0\cos\left(a\omega\right)\right] + \nonumber \\
&& \frac{e^{-b\sigma}}{3r^2}\left[bB^2e^{-b\sigma}\left(3+b\sigma\right)+
bBw_0\cos\left(b\omega\right)\right] + \nonumber \\
&& \frac{ABe^{-\left(a+b\right)\sigma}}{3r^2}\left[
3\left(a+b\right)+2ab\sigma
\right]\cos\left(\left(a-b\right)\omega\right)
\label{racetrack_VF}
\earay
The minimal form of the functions $f_{1}$ and $f_{2}$ brings an important simplification 
as the dependence of the F-term potential on the positions of the mobile branes comes only 
through $r = T+\overline{T}- \phi^{1}\overline{\phi^{1}} - \phi^{2}\overline{\phi^{2}}$.

The anti-branes that lift the minimum of the potential to a deSitter one during inflation
have a contribution of the form:
\begin{equation}
V_{sb} = \sum_{i=1}^{N}\frac{E_{i}}{r^{\alpha}}.
\end{equation}
The coefficients $E_i$ are the warped anti-brane tension. 
The exponent $\alpha = 2$ for anti-branes located in a highly warped
region, i.e. at the bottom of the throat \cite{Kachru:2003sx}. It is
not necessary that all branes are located inside the same throat 
\cite{Iizuka:2004ct}, but we will consider a single-throat model for simplicity.

We also have to add the Newtonian attractive pieces for the branes and anti-branes:
\begin{equation}
V_{N} = -\sum_{i=1, j=1}^{M, N}\frac{k_{ij}}{r^2\left(\phi^{i}-\phi_{0}^{j}\right)^{4}}.
\end{equation}
We want to trace the evolution of more than one brane, and the above potential is
singular when an brane collides with an anti-brane. Therefore we want to
regulate the Newtonian potential such that it reproduces the correct inverse-power-law at large
distances, but at the same time is regular and cancels the
brane-anti-brane tension at zero separation. We follow here the method
of Ref. \cite{Cline:2005ty} and for a given brane-anti-brane pair we choose
\begin{equation}
V_{N}\left(\phi^i\right)=-\sum_{i=1, j=1}^{M, N}
\frac{k_{ij}}{r^2\left[s^{ij}+\left(\phi^{i}-\phi_{0}^{j}\right)^{4}\right]},
\end{equation}
where the constant $s^{ij}$ is chosen such that when $\phi^{i}=\phi_{0}^{j}$, the 
Newtonian potential exactly cancels the contribution to the potential of the anti-brane
placed at $\phi_{0}^{j}$. If there are more than one anti-brane located at $\phi_{0}^{j}$,
we choose $s^{ij}$ to cancel the tension of only one of them.

To summarise, the choice of K\"ahler potential and the assumption that we have a 
single-throat model, such that all anti-branes are located at the same point, allows 
us to write the scalar potential for a model with $n$ mobile branes and $n$ anti-branes as:
\begin{equation}
V\left(\sigma, \phi^{1}\dots\phi^{n}\right) = 
\frac{V_{0}}{r^2}\left[F\left(\sigma\right) + nT - 
\sum_{i=1}^{n}\frac{nk}{s^{i}+\left(\phi^{i}-\phi_{0}\right)^{4}}\right].
\label{full_potential}
\end{equation}
The function $F\left(\sigma\right)$ contributed by the F-term is 
responsible for the stabilisation of the volume modulus at some value 
$\sigma_{0}$, and, as long as we do not add too many 
anti-branes to destabilise this minimum, we can simply approximate 
$F\left(\sigma\right) = \frac{1}{2}m_{\sigma}^{2}\left(\sigma-\sigma_{0}\right)^{2}$ 
when studying the cosmological signatures of the model. 

\section{Properties of the field space metric}
We want to get a better understanding of the properties of the target (field) space.
In order to do so we would like to write the K\"ahler metric for the real
parts of the fields in a more convenient coordinate system. 
\baray
dS^2 &=& \frac{6}{r^2}\left[d\sigma^2 -2 \phi^{i}d\phi^{i}d\sigma + 
rd\phi^{i}d\phi^{i} + \phi^{i}\phi^{j}d\phi^{i}d\phi^{j}\right] = 
\nonumber \\
&& \frac{6}{r^2}\left[\left(d\sigma - \phi^{i}d\phi^{i}\right)^2 + 
rd\phi^{i}d\phi^{i}\right] = \frac{6}{r^2}\left[\frac{1}{4}dr^2 + 
rd\phi^{i}d\phi^{i}\right]
\earay
We can now redefine the coordinate $r$ as $r = u^{2}$. This brings the metric 
in the form:
\begin{equation}
dS^2 = \frac{6}{u^{4}}\left[\frac{4u^2 du^2}{4} + 
u^2d\phi^{i}d\phi^{i}\right] = 
\frac{6}{u^{2}}\left[du^{2} + \sum_{i=1}^{n}{d\phi^{i}}^{2}\right] , 
\end{equation}
which is the metric for the $n+1$ - dimensional Euclidean AdS space. 
The original coordinates cover only a limited patch of the space, 
the boundary of it being given by the condition:
\begin{equation}
2\sigma - \sum_{i=1}^{n}{\phi^{i}}^{2} > 0,
\end{equation} 
which translates into the condition that $u>0$. This is the Poincar\'e
half-space model of AdS,  \cite{Caldarelli:1998wk}. The expressions
for the metric connection coefficients are much simpler in the 
new variable. Up to symmetries in permuting the indices:
\baray
\Gamma^{u}_{~uu} &=& -\frac{1}{u} \nonumber \\
\Gamma^{u}_{~ij} &=& \frac{\delta_{ij}}{u} \nonumber \\
\Gamma^{i}_{~uj} &=& -\frac{\delta_{ij}}{u} \nonumber \\
\earay
All other connection coefficients, most notably $\Gamma^{i}_{~jk}$
vanish and the non-vanishing ones are now independent of the brane
positions. As far as the potential of the effective theory is
concerned, one has to replace $\sigma$ by the new variable $u$:
\begin{equation}
r=2\sigma-\sum_{i=1}^{n}{\phi^{i}}^{2} = u^2, ~~~ 
\sigma = \frac{1}{2}\left(u^2 + \sum_{i=1}^{n}{\phi^{i}}^{2}\right).
\end{equation}
In KKLT type of models the potential has the form Eq.~(\ref{full_potential}), 
this change of variables brings the potential to the form:
\begin{equation}
V\left(u, \phi^{1}\dots\phi^{n}\right) = 
\frac{V_{0}}{u^4}\left[F\left(\frac{u^2}{2} + 
\frac{1}{2}\sum_{i=1}^{n}{\phi^{i}}^{2}\right) 
+ nT - \sum_{i=1}^{n}\frac{nk}{\left(\phi^{i}-\phi_{0}\right)^{4}}\right].
\end{equation}
The volume-stabilising function $F$ preserves the full $SO\left(n\right)$ rotational 
symmetry of the brane-position inflatons, and the dependence on the brane positions is 
removed from the overall factor $1/u^{4}$. Combined with the fact that in terms of the 
new variables the field-space metric is independent of the positions of the branes, 
the Newtonian brane-anti-brane attraction remains the only part of the potential 
which sets a preferred direction in the field space. This will lead to a field trajectory
during slow roll that will be a straight line, and therefore to very small signatures 
in the CMB power spectrum.

\section{Field evolution and inflation} 
To study the dynamics of the fields during inflation we start with the 
action for the scalar fields coupled to gravity:
\begin{equation} 
S = \int d^3xdt\sqrt{-g}\left(-\hf G_{ij}\partial_{\mu}\phi^{i} 
\partial^{\mu}\phi^{j}-V+\frac{R}{16\pi G}\right).
\end{equation} 
We assume that the space-time metric is of the FRW type,  
\begin{equation} 
\label{FRW_metric}
ds^2 = -dt^2+a^2\left(t\right)\left(dr^2+r^2d\Omega^2\right),
\end{equation} 
and we decompose the fields into a background, homogeneous part, 
and a space-dependent perturbation. The equation for the homogeneous 
fields is the geodesic equation in field space with a driving force 
given by the potential gradient and a damping force given  by Hubble 
expansion:   
\begin{equation} 
\label{timeeq} 
\ddot\phi^k + 3H\dot\phi^k +  
G^{ka}\frac{\partial V}{\partial \phi^a} +  
\Gamma_{ij}^{k}\dot\phi^i\dot\phi^j = 0.
\end{equation} 
The Hubble constant and its time derivative are given by: 
\begin{equation} 
H^2 = \frac{8\pi G}{3}\left(\frac{1}{2}G_{ij}\dot\phi^i\dot\phi^j + V\right) 
\;\;,\;\;\;
\dot{H} = -(4\pi G)G_{ij}\dot\phi^i\dot\phi^j.
\end{equation} 
We write here the more general equation for inhomogeneous fields, 
as we will need it later when studying the perturbations around the 
homogeneous background: 
\baray 
\label{inhomogeneous_eq}
&& \frac{1}{\sqrt{-g}}\partial_{\mu} 
\left[\sqrt{-g}g^{\mu\nu}\partial_{\nu}\phi^{K}\right] -  
G^{KA}\frac{\partial V}{\partial \phi^A} +  
\Gamma_{IJ}^{K}g^{\mu\nu}\partial_{\mu}\phi^I\partial_{\nu}\phi^J = 0 \\ 
&& H^2 = \frac{1}{3}\left( 
G_{IJ}g^{\mu\nu}\partial_{\mu}\phi^I\partial_{\nu}\phi^J + V\right) 
\earay 
 
\section{Perturbation equations} 
In this section we analyze the perturbations around the background, slow-roll  
evolution. First, the background fields and metric are only dependent on time,  
while the perturbations are dependent on space as well. We will follow  
Ref.\cite{Gordon:2000hv} and parametrise the metric perturbations as follows: 
\begin{equation} 
g_{\mu\nu} + \delta g_{\mu\nu} =  
\left(\begin{array}{cc} 
-\left(1+2A\left(t, {\bf x}\right)\right) &  
a\left(t\right)B_{, i} \left(t, {\bf x}\right) \\ 
a\left(t\right)B_{, i} \left(t, {\bf x}\right) &  
a^2\left(t\right)\left[\left(1-2\psi\left(t, {\bf x}\right)\right)\delta_{ij}+ 
2E_{,ij}\left(t, {\bf x}\right)\right],
\end{array}\right) 
\label{metric_pert}
\end{equation} 
and that of the fields $\phi^{I}$ as: 
\begin{equation} 
\label{scal_pert}
\phi^{I} = \phi^{I}\left(t\right) + \delta\phi^{I}\left(t, {\bf x}\right).
\end{equation} 
We now expand the evolution equation (\ref{timeeq}) for the fields 
$\phi^{I}$ around the background. We will look at each term in the 
equation separately. 
 
\subsection{Kinetic Term} 
 We start with the first term in Eq.(\ref{inhomogeneous_eq}) and
 write the scalars and the metric as in
 Eq.(\ref{metric_pert},\ref{scal_pert}). The result for the plane wave
 expanded perturbations 
(see appendix \ref{detail_pert_kinetic} for details) is: 
\baray 
&& \delta \left(\frac{1}{\sqrt{-g}}\partial_{\mu} 
\left[\sqrt{-g}g^{\mu\nu}\partial_{\nu}\phi^{K}\right]\right) 
= -\ddot{\delta\phi^{K}} -  
3\frac{\dot{a}}{a}\dot{\delta\phi^{K}} -  
\frac{k^2}{a^2}\delta\phi^{K} 
\nonumber \\ 
&& + 2A\left[\ddot{\phi^{K}} +  
3\frac{\dot{a}}{a}\dot{\phi^{K}}+ 
\Gamma_{IJ}^{K}\dot{\phi^I}\dot{\phi^J} 
\right] +2\dot{A}\dot{\phi^K} -  
\frac{k^2}{a}B\dot{\phi^K} 
-\left[\dot{A}-3\dot{\psi}-k^2\dot{E}\right] \dot{\phi^{K}}, 
\label{pert_kinetic}
\earay
where $k\equiv |{\vec k}|$ is the expansion wavenumber.  
 
\subsection{Potential and Christoffel  Terms} 
 
The perturbation of the potential term is the simplest one, since it does not  
involve the space-time metric. We simply take the variation with respect to the  
fields $\phi^{I}$ : 
\begin{equation} 
\delta\left(G^{KI}\frac{\partial V}{\partial\phi^{I}}\right) =  
\left[G^{KI}\frac{\partial^2 V}{\partial\phi^{I}\partial\phi^{J}} +  
\frac{\partial G^{KI}}{\partial\phi^{J}}\frac{\partial V}{\partial\phi^{I}}\right] 
\delta\phi^{J}.
\end{equation} 
Finally, we have to calculate the perturbation of the connection coefficients 
coming from the non-trivial field-space metric. They are: 
\baray 
&& \Gamma_{IJ}^{K}\left(\phi+\delta\phi\right) 
g^{\mu\nu}\partial_{\mu}\left(\phi^I+\delta\phi^I\right) 
\partial_{\nu}\left(\phi^J+\delta\phi^J\right) =  
\underbrace{\Gamma_{IJ}^{K}g^{\mu\nu}\partial_{\mu}\phi^I\partial_{\nu}\phi^J 
}_{\text{background}} 
\nonumber \\ 
&& + 2\Gamma_{IJ}^{K}g^{\mu\nu}\partial_{\mu}\phi^I\partial_{\nu}\delta\phi^J 
+ \frac{\partial \Gamma_{IJ}^{K}}{\partial \phi^L}  
g^{\mu\nu}\partial_{\mu}\phi^I\partial_{\nu}\phi^J\delta\phi^{L} 
+ \Gamma_{IJ}^{K}\delta g^{\mu\nu}\partial_{\mu}\phi^I\partial_{\nu}\phi^J =
\nonumber \\ 
&&  \underbrace{\Gamma_{IJ}^{K}g^{\mu\nu}\partial_{\mu}\phi^I\partial_{\nu}\phi^J 
}_{\text{background}} 
-2\Gamma_{IJ}^{K}\dot{\phi^I}\dot{\delta\phi^J} 
-\frac{\partial \Gamma_{IJ}^{K}}{\partial \phi^L}  
\dot{\phi^I}\dot{\phi^J}\delta\phi^{L} + 
2A\Gamma_{IJ}^{K}\dot{\phi^I}\dot{\phi^J} 
\earay 
 We can now collect the perturbation terms and write the equation for the  
fluctuations of the fields: 
\baray 
&& -\ddot{\delta\phi^{K}} -  
3\frac{\dot{a}}{a}\dot{\delta\phi^{K}} -  
\frac{k^2}{a^2}\delta\phi^{K} +  
2A\underbrace{\left[\ddot{\phi^{K}} +  
3\frac{\dot{a}}{a}\dot{\phi^{K}}+ 
\Gamma_{IJ}^{K}\dot{\phi^I}\dot{\phi^J} 
\right]}_{=-2AG^{KI}\frac{\partial V}{\partial\phi^{I}}}  
+2\dot{A}\dot{\phi^K} -  
\frac{k^2}{a}B\dot{\phi^K} 
\nonumber \\ 
&& -\left[\dot{A}-3\dot{\psi}-k^2\dot{E}\right] \dot{\phi^{K}} -  
\left[G^{KI}\frac{\partial^2 V}{\partial\phi^{I}\partial\phi^{J}} +  
\frac{\partial G^{KI}}{\partial\phi^{J}}\frac{\partial V}{\partial\phi^{I}}\right] 
\delta\phi^{J} 
\nonumber \\ 
&& -2\Gamma_{IJ}^{K}\dot{\phi^I}\dot{\delta\phi^J} 
-\frac{\partial \Gamma_{IJ}^{K}}{\partial \phi^L}  
\dot{\phi^I}\dot{\phi^J}\delta\phi^{L} = 0 
\earay 
Using the equation for the background fields, the perturbation 
equation can be rewritten as: 
\baray 
&& \ddot{\delta\phi^{K}}  + 
3\frac{\dot{a}}{a}\dot{\delta\phi^{K}} +  
\frac{k^2}{a^2}\delta\phi^{K} +  
\left[G^{KI}\frac{\partial^2 V}{\partial\phi^{I}\partial\phi^{J}} +  
\frac{\partial G^{KI}}{\partial\phi^{J}}\frac{\partial V}{\partial\phi^{I}}\right] 
\delta\phi^{J} 
\nonumber \\ 
&& = -2AG^{KI}\frac{\partial V}{\partial\phi^{I}} 
+ \left[\dot{A} + 3\dot{\psi} + k^2\dot{E}  -  
\frac{k^2}{a}B\right]\dot{\phi^{K}} 
\nonumber \\ 
&& -2\Gamma_{IJ}^{K}\dot{\phi^I}\dot{\delta\phi^J} 
-\frac{\partial \Gamma_{IJ}^{K}}{\partial \phi^L}  
\dot{\phi^I}\dot{\phi^J}\delta\phi^{L} = 0 
\label{pertEquns} 
\earay 
 
\section{Metric Perturbations} 
 
In this section we analyse the perturbations of the Einstein equations,  
namely the perturbations of the Einstein tensor and stress-energy tensor.  
We start with the parametrisation of the metric perturbations given in 
Eq. (\ref{metric_pert}): 
\begin{equation} 
\delta g_{\mu\nu}= 
\left(\begin{array}{cccc} 
-2A & aB_{,1} & aB_{,2} & aB_{,3} \\ 
aB_{,1} & 2a^2\left(-2\psi+E_{,11}\right) & 2a^2E_{,12} & 2a^2E_{,13} \\ 
aB_{,2} & 2a^2E_{,21} & 2a^2\left(-2\psi+E_{,22}\right) & 2a^2E_{,23} \\ 
aB_{,3} & 2a^2E_{,31} & 2a^2E_{,32} & 2a^2\left(-2\psi+E_{,33}\right) \\ 
\end{array}\right).
\end{equation}  
The functions $A$, $\psi$, $B$, and $E$ are in general dependent upon 
all coordinates, $t, x, y, z$. We follow here Ref. \cite{Mukhanov:1990me} 
and perturb the Einstein equations written in the form: 
\begin{equation} 
R_{\mu}^{\nu} - \hf R \delta_{\mu}^{\nu} = 8\pi G T_{\mu}^{\nu}. 
\label{Einstein_background}
\end{equation} 
First, the perturbations of the Einstein tensor is: 
\baray 
\delta\left(R_{0}^{0} - \hf R \delta_{0}^{0}\right) &=&  
6\frac{\dot a}{a}\left(\frac{\dot a}{a}A + \dot{\psi}\right) 
+2\frac{k^2}{a^2}\left[\psi+\frac{\dot a}{a}\left(a^2\dot{E}-aB\right) 
\right]\\ 
\delta R_{0}^{i} &=& 2\left(\dot{\psi}+\frac{\dot a}{a}A\right)_{,i} 
\earay 
For the stress-energy tensor the general expression is: 
\begin{equation} 
T_{\alpha}^{\beta}=- 
\left[\hf G_{IJ}g^{\mu\nu}\partial_{\mu}\phi^{I}\partial_{\nu}\phi^{J} 
+V\right]\delta_{\alpha}^{\beta}+ 
G_{IJ}\partial_{\alpha}\phi^{I}\partial^{\beta}\phi^{J}. 
\end{equation} 
We now perturb the stress-energy tensor according to our parametrisation of 
the metric and field perturbations we obtain: 
\baray 
&& \delta T_{\alpha}^{\beta} = -\left[ 
\hf G_{IJ}\delta g^{\mu\nu}\partial_{\mu}\phi^{I}\partial_{\nu}\phi^{J} 
+ G_{IJ}g^{\mu\nu}\partial_{\mu}\phi^{I}\partial_{\nu}\delta \phi^{J} 
\right. 
\nonumber \\ 
&& \left.+ \hf \frac{\partial G_{IJ}}{\partial\phi^{K}} 
g^{\mu\nu}\partial_{\mu}\phi^{I}\partial_{\nu}\phi^{J}\delta\phi^{K} 
+\frac{\partial V}{\partial\phi^{K}}\delta\phi^{K} 
\right]\delta_{\alpha}^{\beta} 
\nonumber \\ 
&& +G_{IJ}(\partial_{\alpha}\delta\phi^{I}\partial_{\gamma}\phi^{J}+ 
\partial_{\alpha}\phi^{I}\partial_{\gamma}\delta\phi^{J})g^{\gamma\beta}
\nonumber \\ 
&& + \frac{G_{IJ}}{\partial \phi^{K}}\partial_{\alpha}\phi^{I}\partial^{\beta}\phi^{J} 
\delta\phi^{K}+ 
G_{IJ}\partial_{\alpha}\phi^{I}\partial_{\gamma}\phi^{J}\delta g^{\gamma\beta} 
\label{SET_Pert} 
\earay 
We can use here the fact that the background fields are homogeneous 
to further simplify the expression for $\delta T_{0}^{0}$. We obtain: 
\baray 
&& \delta T_{0}^{0} =  
\left[-\hf G_{IJ}\delta g^{00}\dot{\phi^{I}}\dot{\phi^{J}}+ 
G_{IJ}\dot{\phi^{I}}\dot{\delta\phi^{J}}+ 
\hf \frac{\partial G_{IJ}}{\partial\phi^{K}} 
\dot{\phi^{I}}\dot{\phi^{J}}\delta\phi^{K} 
-\frac{\partial V}{\partial\phi^{K}}\delta\phi^{K}\right]\delta_{0}^{0} 
\nonumber \\ 
&& -\frac{\partial G_{IJ}}{\partial\phi^{K}} 
\dot{\phi^{I}}\dot{\phi^{J}}\delta\phi^{K} 
-2G_{IJ}\dot{\phi^{I}}\dot{\delta\phi^{J}} 
+G_{IJ}\dot{\phi^{I}}\dot{\phi^{J}}\delta g^{00} =  
\nonumber \\ 
&& AG_{IJ}\dot{\phi^{I}}\dot{\phi^{J}} -  
G_{IJ}\dot{\phi^{I}}\dot{\delta\phi^{J}} -  
\hf \frac{\partial G_{IJ}}{\partial\phi^{K}} 
\dot{\phi^{I}}\dot{\phi^{J}}\delta\phi^{K} 
-\frac{\partial V}{\partial\phi^{K}}\delta\phi^{K} 
\earay 
For the component $T_{0}^{i}$ the background value is trivial, 
$T_{0}^{i} = 0$ and the only term that contributes is the one in the   
next-to-last line in Eq.(\ref{SET_Pert}): 
\begin{equation} 
\delta T_{0}^{i} = G_{IJ}\dot{\phi^{I}}\partial_{i}\delta\phi^{J}. 
\end{equation} 
Therefore the equation for the evolution of the metric perturbations become: 
\baray 
&& 2\left(\dot{\psi}+\frac{\dot a}{a}A\right)_{,i} =  
8\pi G \cdot G_{IJ}\dot{\phi^{I}}\delta\phi^{J}_{,i} \label{momentum} 
\\ 
&& 6\frac{\dot a}{a}\left(\frac{\dot a}{a}A + \dot{\psi}\right) 
+2\frac{k^2}{a^2}\left[\psi+\frac{\dot a}{a}\left(a^2\dot{E}-aB\right) 
\right]=  
\nonumber \\ 
&& 8\pi G \left[AG_{IJ}\dot{\phi^{I}}\dot{\phi^{J}} -  
G_{IJ}\dot{\phi^{I}}\dot{\delta\phi^{J}} -  
\hf \frac{\partial G_{IJ}}{\partial\phi^{K}} 
\dot{\phi^{I}}\dot{\phi^{J}}\delta\phi^{K} 
-\frac{\partial V}{\partial\phi^{K}}\delta\phi^{K}\right] \label{energy} 
\earay 
\section{The evolution and spectrum of the perturbations}
We now proceed to deriving the equation of motion for the gauge-invariant 
Mukhanov-Sasaki variables
\begin{equation}
Q^{I}=\delta\phi^{I}+\frac{\dot{\phi^{I}}}{H}\psi.
\end{equation}
For the single field, slow-roll case the equations of motions for the
variable $a\,Q$ assume a particularly simple form of an oscillator with a
time (background only) dependent mass. The perturbations in the scalar
sector are then easily integrated from an initial state which can be
approximated by a Minkowski adiabatic vacuum for $k \gg aH$ to a final
state at horizon crossing $k=aH$ after which the perturbation freezes
out and stops evolving. 

For multiple field models two complications can arise. The first is
that mixing terms in the multiple field potential will lead to a
system of coupled oscillators with a non-diagonal mass matrix. The
second is that a non-trivial field-space metric (kinetic terms) leads
to the presence of connection terms in the equations. Another
important difference with the single field case is that, in general,
the perturbations will continue evolving once they cross the horizon
and must be solved for well into the super-horizon regime.

In what follows we use the spatially flat gauge in which  
\baray 
\psi_Q & = & 0,  
\\ 
a^2\dot E_Q - aB_Q & = & \frac{1}{H} \Psi,
\earay 
where the subscript $Q$ denotes the use of the spatially flat gauge 
and $\Psi$ is the curvature perturbation. This gauge makes the equations 
simple as the Mukhanov-Sasaki variables reduce to the simple form: 
\begin{equation} 
Q^I = \delta\phi^I. 
\end{equation} 
Introducing the choice of gauge into Eq.~(\ref{momentum}) leads the to
following expression for $A_Q$: 
\begin{equation} 
A_Q = \left(\frac{4\pi G}{H}\right) G_{IJ}\dot{\phi^{I}}\delta\phi^{J}. 
\end{equation} 
Differentiating this we then obtain 
\begin{equation} 
\dot A_Q = \left(\frac{4\pi G}{H}\right)\left[
-\frac{\dot H}{H} G_{IJ}\dot\phi^I\delta\phi^J 
+\frac{\partial G_{IJ}}{\partial\phi^{K}}\dot{\phi^{I}}\delta{\phi^{J}}\dot\phi^{K} 
+G_IJ\ddot\phi^I\delta\phi^J  
+G_IJ\dot\phi^I\dot{\delta\phi^J }\right].
\label{Adot} 
\end{equation} 
The value for $\Psi$ comes from Equation~(\ref{energy}): 
\baray 
3H^2A_Q + \frac{k^2}{a^2}\Psi = 4\pi G \left[A_QG_{IJ}\dot{\phi^{I}}\dot{\phi^{J}} -  
G_{IJ}\dot{\phi^{I}}\dot{\delta\phi^{J}} -  
\hf \frac{\partial G_{IJ}}{\partial\phi^{K}} 
\dot{\phi^{I}}\dot{\phi^{J}}\delta\phi^{K} 
-\frac{\partial V}{\partial\phi^{K}}\delta\phi^{K}\right].
\earay 
Thus the curvature perturbation has the expression: 
\baray 
\frac{1}{H}\frac{k^2}{a^2}\Psi & = &  
\left(\frac{4\pi G}{H}\right)\left[ - 3HG_{IJ}\dot\phi^I\delta\phi^J+ 
\left(\frac{4\pi G}{H}\right) 
G_{LM}\dot{\phi^{L}}\dot{\phi^{M}}G_{IJ}\dot{\phi^{I}}\delta\phi^J \right.
\nonumber \\ 
& & \left.- G_{IJ}\dot{\phi^{I}}\dot{\delta\phi^{J}} - 
\hf \frac{\partial G_{IJ}}{\partial\phi^{K}}\dot{\phi^{I}}\dot{\phi^{J}}\delta\phi^{K} - 
\frac{\partial V}{\partial\phi^{K}}\right]\delta\phi^{K}.
\label{Psi} 
\earay

Using the previously-derived expressions for $A_{Q}$ and $\Phi$ we can
now calculate 
the gauge terms in the right-hand-side of the equations of motion of the 
perturbation variables, Equation (\ref{pertEquns}). 
\baray 
-2AG^{KI}\frac{\partial V}{\partial\phi^{I}} & + & 
\left[\dot{A} + 3\dot{\psi} + k^2\dot{E} - \frac{k^2}{a}B\right]\dot{\phi^{K}}  =  
-2A_QG^{KI}\frac{\partial V}{\partial\phi^{I}} + \left[\dot{A_Q} + 
\frac{1}{H}\frac{k^2}{a^2}\Psi \right]\dot{\phi^{K}} 
\nonumber \\ 
& = & \left(\frac{8 \pi G}{H}\right)\left[G_{IJ}\ddot\phi^K\dot\phi^I\delta\phi^J+
3HG_{IJ}\dot\phi^K\dot\phi^I\delta\phi^J+
\Gamma_{ILJ}\dot\phi^K\dot\phi^I\dot\phi^L\delta\phi^J \right. 
\nonumber \\ 
&   & \left.+ G_{IJ}\dot\phi^K\ddot\phi^I\delta\phi^J -
\frac{\dot H}{H} G_{IJ}\dot\phi^K\dot\phi^I\delta\phi^J\right] 
\nonumber \\ 
& = & \left[\left(\frac{8 \pi G}{a^3}\right) 
G_{IJ}\frac{\partial}{\partial t} \left(\frac{a^3\dot\phi^K\dot\phi^I}{H}\right)+
\left(\frac{8 \pi
  G}{H}\right)\Gamma_{ILJ}\dot\phi^K\dot\phi^I\dot\phi^L
\right]\delta\phi^J 
\label{A_Q}
\earay 
 
Now we combine all the above results, 
Equations~(\ref{Adot}), (\ref{Psi}), and (\ref{A_Q}) can now be
combined to rewrite the equation of motion for each perturbation
$\delta \phi^I$, Eq.~(\ref{pertEquns}), in terms of the new variable $Q^I$
\baray \label{eq:Q}
&&\ddot{Q^{K}} + 3H\dot{Q^{K}} +\frac{k^2}{a^2}Q^{K}  
+2\Gamma_{IJ}^{K}\dot{\phi^I}\dot{Q^J}  
+\left[G^{KI}\frac{\partial^2 V}{\partial\phi^{I}\partial\phi^{J}}  
\right. 
\nonumber\\  
&&+  \frac{\partial G^{KI}}{\partial\phi^{J}}\frac{\partial V}{\partial\phi^{I}} 
+ \frac{\partial \Gamma_{IJ}^{K}}{\partial \phi^L}\dot{\phi^I}\dot{\phi^J}\dot\phi^{L} 
-\left(\frac{8 \pi G}{a^3}\right)  
G_{IJ}\frac{\partial}{\partial t} \left(\frac{a^3\dot\phi^K\dot\phi^I}{H}\right) 
\nonumber\\  
&& \left.
-\left(\frac{8 \pi G}{H}\right)\Gamma_{ILJ}\dot\phi^K\dot\phi^I\dot\phi^L 
-\left(\frac{8 \pi G}{H}\right)\Gamma^{K}_{LM}\dot\phi^{L}\dot\phi^{M}G_{IJ}\dot\phi^{I} 
\right]Q^{J} = 0.
\earay 
Notice that all coefficients in the equation above depend 
only on background quantities. For a canonical kinetic term, 
$G_{IJ} = \delta_{ij}$, all the connection coefficients vanish
and the only mixing of the Mukhanov-Sasaki variables $Q^{K}$ comes
from the mass matrix of the inflatons. The mixing provided by the 
non-trivial field-space metric enhances the possibility that 
non-trivial features on the power spectrum can be detectable.
 
\subsection{Adiabatic and Entropy Perturbations} 

Equation~(\ref{eq:Q}) allows one to integrate efficiently the system
of perturbations from an initial state well inside the
horizon to a final, super-horizon state at the end of inflation when
the perturbations determine initial conditions in the constituents of
the universe after a period of preheating. 

In the single field scenario the perturbations in the inflaton are
simply related to perturbations in the curvature. If multiple fields
are present however there can exist entropy perturbations between the
different fields which leave the curvature unperturbed. The two types
of perturbations induce different initial conditions for the fluid
perturbations at the beginning of the radiation era and lead to
different CMB signatures \cite{cmbisocurv}.

In terms of the generalized trajectory for the variables $Q^K$ one can
define a single adiabatic (or curvature) and $n-1$ entropy
perturbations at any point along the trajectory. The decomposition
allows one to identify the observationally distinct perturbations at
the end of inflation. The variables $Q^{K}$ define a ``perturbation vector'', $\ve{Q}$ in
field space. We can decompose this vector in two components: the one
parallel to the velocity of the background fields and the component
orthogonal to it. We start by defining the unit field velocity vector,
or the adiabatic direction. 
\begin{equation} 
\ve{a} = \frac{\dot{\phi^i}}{\sqrt{G_{kl}\dot{\phi}^k\dot{\phi}^l}}.
\end{equation} 
The adiabatic perturbation is defined by projecting the perturbation 
vector $\ve{Q}$ along the adiabatic direction $\ve{a}$.
\begin{equation} 
Q_{\sigma} =
\frac{G_{ij}\dot{\phi^i}Q^j}{\sqrt{G_{kl}\dot{\phi}^k\dot{\phi}^l}}=
\frac{\dot{\phi^i}\delta\phi^{j}}{\sqrt{G_{kl}\dot{\phi}^k\dot{\phi}^l}} 
+\frac{\sqrt{G_{kl}\dot{\phi}^k\dot{\phi}^l}}{H}\psi.
\end{equation}\label{eq:Qsig}
The corresponding entropy direction is defined as the component of
$\ve{Q}$ orthogonal to $\ve{a}$:
\begin{equation}
\ve{e} =\frac{\ve{Q}-\left(\ve{a}\cdot\ve{Q}\right)\ve{a}}
{\left|\ve{Q}-(\ve{a}\cdot\ve{Q})\ve{a} \right|}.
\end{equation} 
The entropy perturbation is the projection of $\ve{Q}$ along $\ve{e}$:
\begin{equation}
Q_{s} = \frac{\left|\ve{Q}\right|^{2} - 
\left(\ve{a}\cdot\ve{Q}\right)^{2}}
{\left|\ve{Q}-(\ve{a}\cdot\ve{Q})\ve{a} \right|}.
\end{equation}\label{eq:Qs}
We can now calculate the standard quantities, the co-moving curvature 
perturbation ${\mathcal R}$ and the re-normalised entropy perturbation,
${\mathcal S}$. 
\begin{eqnarray} 
{\mathcal R} & = & \frac{H}{\sqrt{G_{ij}\dot\phi^i\dot\phi^j}}Q_\sigma \\ 
{\mathcal S} & = & \frac{H}{\sqrt{G_{ij}\dot\phi^i\dot\phi^j}}Q_{s} 
\end{eqnarray} 
The corresponding power spectra for the two variables are:
\baray
P_{\mathcal R}\left({\bf k}\right) &=&
\frac{k^3}{2\pi^2}\left|{\mathcal R}\left(k\right)\right|^2,  \\
P_{\mathcal S}\left({\bf k}\right) &=&
\frac{k^3}{2\pi^2}\left|{\mathcal S}\left(k\right)\right|^2.
\earay
These are the two quantities that we plot for a number of
configurations. We look at a single brane model, as well as
multi-brane models (3 mobile branes) in which the branes start
either from the same location, or from slightly different positions. 
We find that the spectrum of the adiabatic perturbation is 
insensitive to the initial conditions and it has the same shape for
both single-brane and multi-brane configurations. Its shape also
agrees with the calculations done in Ref. \cite{Burgess:2004kv,Cline:2005ty}
using the Sasaki-Stewart formula \cite{Sasaki:1995aw}.

\section{Numerical solution of power spectra}

\begin{figure}[thbp]
\begin{center}
\includegraphics[width=7cm,angle=0]{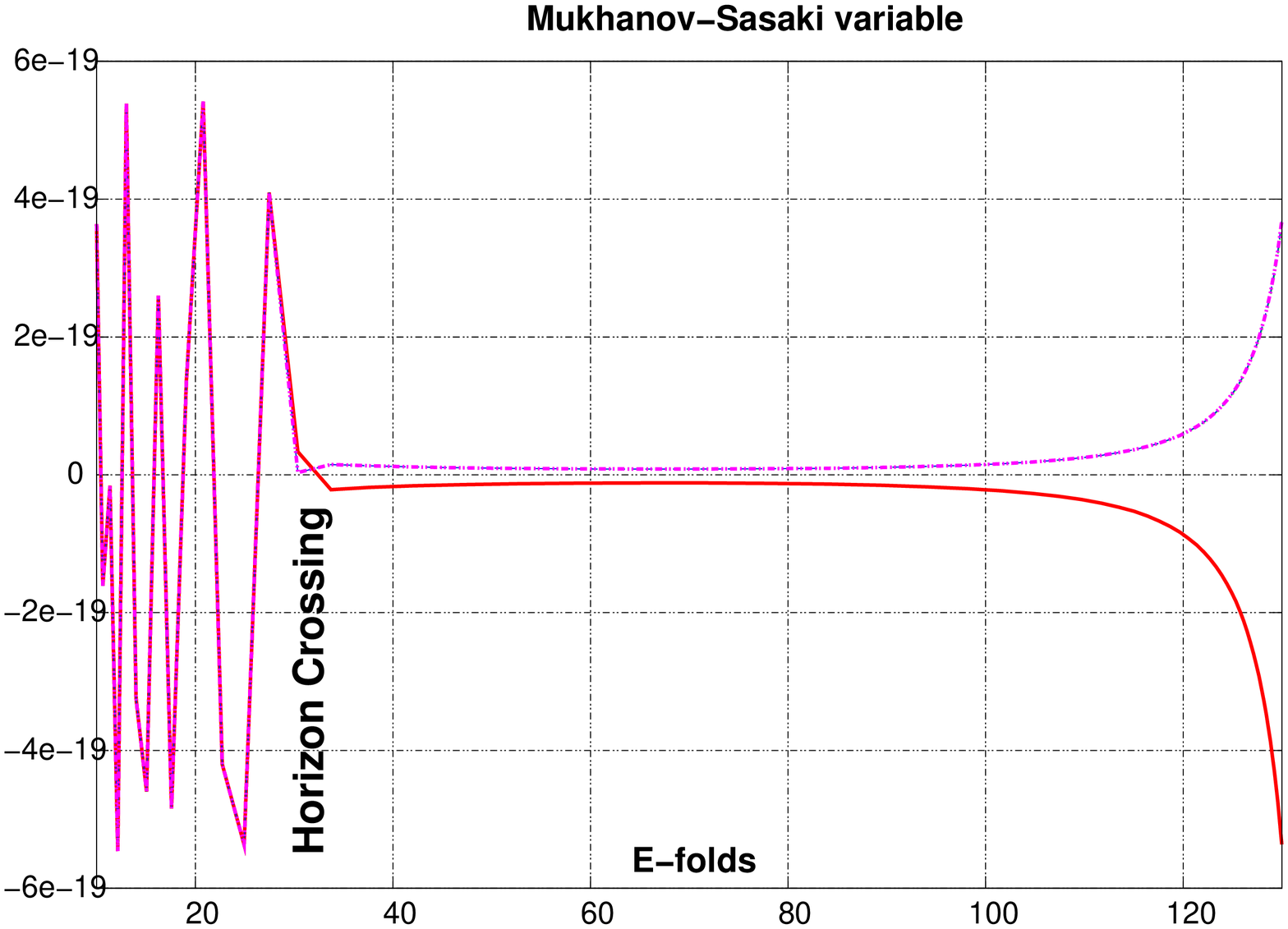}
\hspace{12pt}
\includegraphics[width=7cm,angle=0]{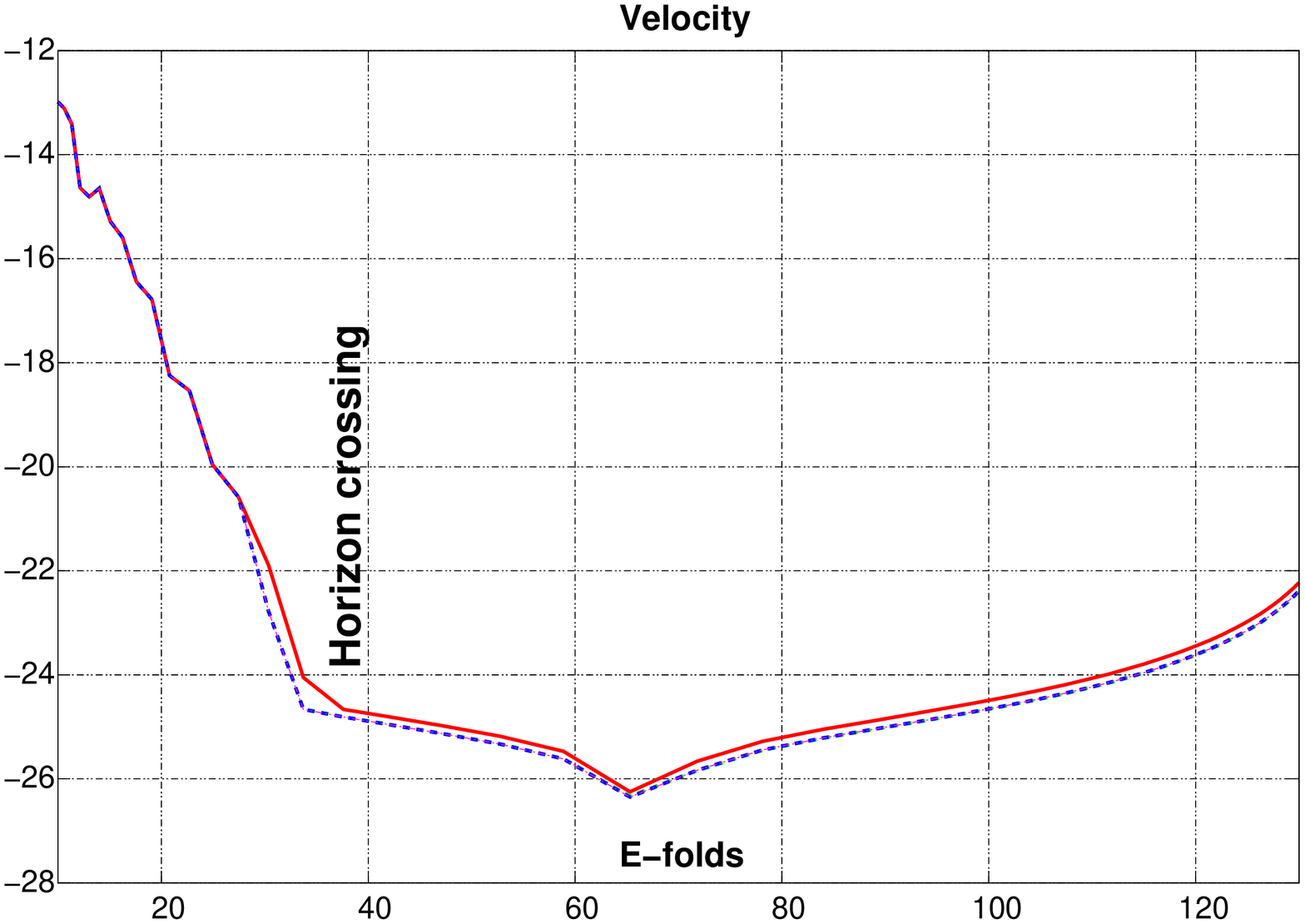}
\caption{Left: Evolution of the Mukhanov-Sasaki variable. The non-trivial shape of the 
  potential leads to evolution of the variable after horizon crossing. The late
  time amplification is due to the change in curvature of the potential after the 
  inflection point; the negative mass squared of the inflaton leads to a tachyonic
  amplification of the modes. 
  Right: The corresponding time derivative of the same Mukhanov-Sasaki variable. We plot
  the absolute value of the velocity on a log scale. The velocity increases after the 
  inflaton passes the inflection point. 
  To calculate the adiabatic and entropy spectra we take a snapshot just before inflation ends.
\label{MS_var}}
\end{center}
\end{figure}

To qualitatively study the spectrum of the perturbations produced in
these models we integrate numerically the evolution of 100
perturbation modes in a model with 3 mobile branes.  We take an
exponential sampling of the momenta such that we have a uniform
sampling as a function of the number of $e$-folds throughout the entire
inflationary period. The modes are evolved from a time 3 $e$-folds
before horizon crossing. Before that the evolution can be solved
analytically under the assumption that the mode is evolving deep
inside the horizon in a deSitter background. The integration is
continued until the background stops inflating. Fig.~\ref{MS_var}
shows the evolution of the Mukhanov-Sasaki variable for a single
wavenumber together with its velocity. The perturbation is initially
constant once it crosses the horizon but then starts growing due to
the roll-over after the inflection point.

A snapshot of the perturbations is taken just before inflation ends and
these are then decomposed into adiabatic and entropic components
following Eqs.~(\ref{eq:Qsig}) and (\ref{eq:Qs}) to obtain the final
super-horizon power spectra.  

\begin{figure}[thbp]
\begin{center}
\includegraphics[width=12cm,angle=270]{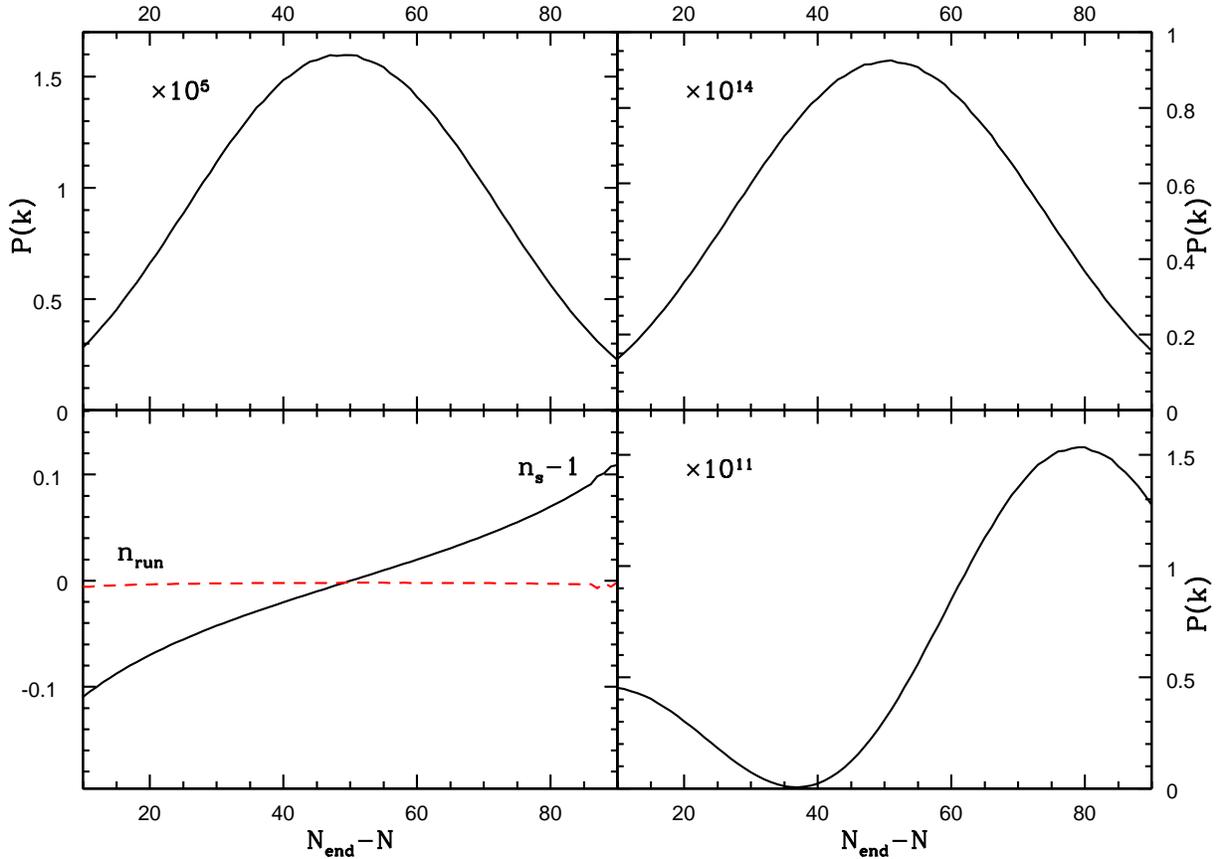}
\caption{Upper Left: The adiabatic spectrum has the same shape for
  single or multi-brane models and is insensitive to the initial
  positions of the branes. Upper Right: The entropy spectrum for
  coincident branes or a single brane. Lower Right: The entropy
  spectrum for the same model, with the branes starting from different
  locations. The amplitude of the spectrum is still 6 orders of
  magnitude smaller than that of the adiabatic perturbations
  Lower Left: The spectral index and its running for the adiabatic
  perturbations. The spectrum is blue before the inflaton passes the
  inflection point, so to get a red spectrum as suggested by the WMAP 
  data we need 60 $e$-folds of inflation after passing the inflection point.
\label{coincident_branes}}
\end{center}
\end{figure}

The single-brane model give the same qualitative results as the
multi-brane model in which all the brane are initially coincident, 
and then roll together.  We choose the parameters such 
that the adiabatic perturbations are COBE normalized for both
the single-brane and the multi-brane models. The amplitude of the
entropy  spectrum is smaller for the single-brane case, but the spectra
have the same shape. The results are shown in
Fig.~\ref{coincident_branes} together with the spectral index $n_s$ and
running $dn_s/d\ln k$ as a function of $e$-folds. 

The spectrum of adiabatic perturbations is initially blue before the
inflection point and then becomes red. The running of the spectral
index is small throughout the trajectory. 

The shape of the entropy spectrum is highly sensitive to
the trajectory in field space, a small change in the initial positions
of the mobile branes results in a completely different shape of the
spectrum as compared with the case of coincident branes. However the
amplitude of the entropy spectrum is 6 orders of magnitude (9 orders
of magnitude for coincident branes) smaller than the amplitude of the
adiabatic perturbations. Such a small contribution to an isocurvature
component of the primordial perturbations would be unobservable be
even the most sensitive future CMB experiments.

\begin{figure}[thbp]
\begin{center}
\includegraphics[width=12cm]{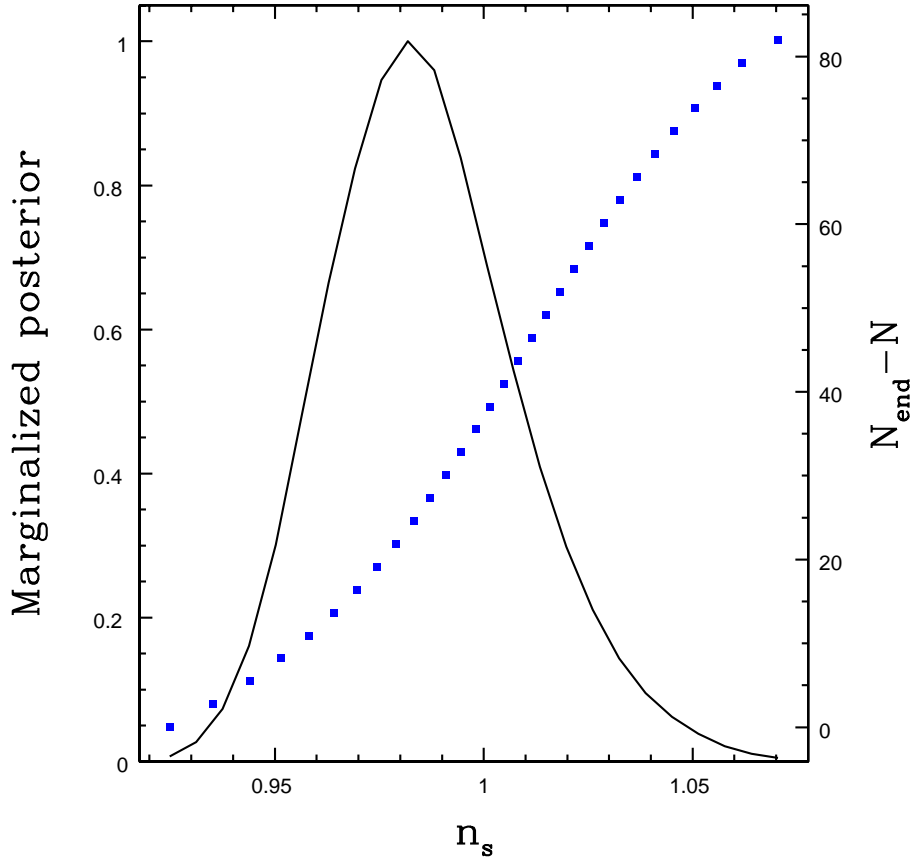}
\caption{The marginalized posterior for the spectral index from the
  ``WMAP5yr + CMB'' \cite{Dunkley} combination for the case where
  running of the spectral index is assumed to be zero. The square
  points are the value of the spectral tilt of the adiabatic
  perturbations shown in Fig.~\ref{coincident_branes} which crossed the
  horizon at different times during inflation (shown on left
  axis). Only perturbations crossing the horizon at late times are
  compatible with the red spectra favoured by CMB data. 
\label{ns}}
\end{center}
\end{figure}

In Fig.~\ref{ns} we compare the marginalized posterior distribution
for the spectral index obtained from the ``WMAP5yr + CMB'' chains
\footnote{Chains at http://lambda.gsfc.nasa.gov} (see \cite{Dunkley}
for details of the data combination) with the spectral index of the
perturbations which crossed the horizon at different $e$-foldings from
the end of inflation ($N_{\rm end}$). The chains were obtained for a pure
power law model of adiabatic primordial perturbations. Only the modes
which exited the horizon after the inflection point are compatible
with the red spectrum preferred by the data.

\begin{figure}[htbp]
\begin{center}
\includegraphics[width=15cm]{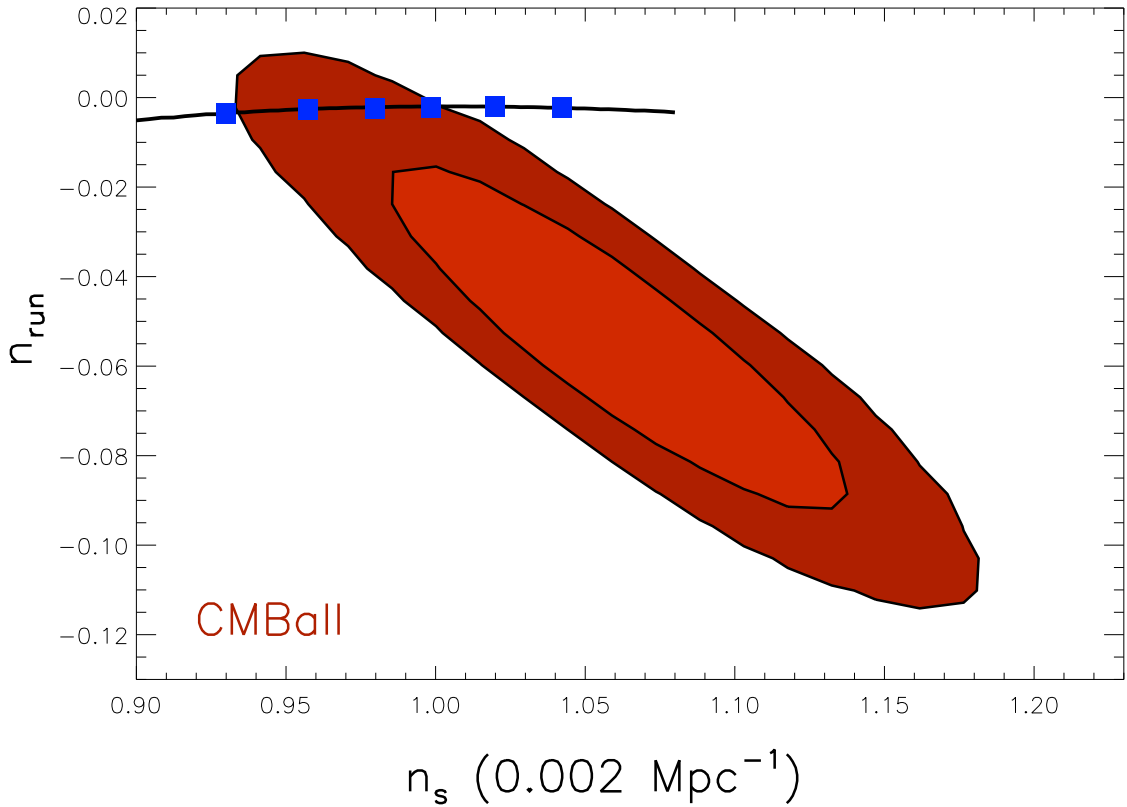}
\caption{The adiabatic perturbations results of Fig.~\ref{coincident_branes}
  plotted over the CMB constraints when both $n_s$ and  $dn_s/d\ln k$
  are fitted. The 68\% and 95\% confidence regions are obtained
  from the ``WMAP5yr + CMB'' \cite{Dunkley}. The
  blue squares on the trajectory indicated, from left to right, the
  point at 20, 30, 40, 50, 60, and 70 $e$-foldings from the 
  end of inflation. The
  smallness of the spectral running in the multi-brane models leads to
  some tension with the observations which prefer more negative
  running. 
  \label{data}}
\end{center}
\end{figure}

The comparison for when running of the spectral, $n_{\rm run}=d
n_s/d\ln k$ index around a pivot point is allowed is shown in
Fig.~\ref{data} for the same data combination. The pivot is at
$k=0.002$ Mpc$^{-1}$ scales probed by current CMB experiments and the
marginalized confidence contours are shown for the $n_s$, $n_{\rm
  run}$ plane. The smallness of the running observed in the models is
in slight disagreement with the data which prefers a negative running
of the index at around the 2-$\sigma$ level. The shift in the
preferred $n_s$ value to $n_s\approx 1.05$ is due to the scale
dependence of the spectral tilt in the presence of non-zero curvature.

\section{Conclusions}

We analysed the possible signatures that brane-inflation models could
give, that may be detectable in observations of the CMB spectrum. To
do so we derived the evolution equation for the perturbations of the
inflatons and metric and numerically evolved a large number of modes
starting inside the horizon, all the way to the end of inflation.  We
then separated the fluctuations into adiabatic and entropy ones and
derived their respective spectra. We found that the adiabatic spectrum
is insensitive to the inflaton trajectories and has the has the same
shape as was determined in earlier work
\cite{Burgess:2004kv,Cline:2005ty} using the Sasaki-Stewart formalism
\cite{Sasaki:1995aw}. The entropy one is highly sensitive to the
inflaton trajectories but its amplitude is too small to be
observable. We also offered an analytical argument that in a large
class of models where most of the $e$-folds come from a small region
around an inflection point of the potential, the inflaton trajectory
is almost a straight line and therefore the entropy perturbations are
expected to be small.  Thus multi-brane models solve a number of
fine-tuning problems but due to their $e$-foldings coming from a reduced
section of the field trajectory they display less phenomenology than
single field models. This argument would lead us to expect that other
potentially observable effects such as non-Gaussianities would also
have a low amplitude in these models. Typically large values of the
non-Gaussian $f_{\rm NL}$ parameter are produced during periods of
high acceleration in the field trajectory which are not present in
these models \cite{ng}.

Interestingly, the old models of brane inflation using
branes oriented at angles and toroidal compactifications
\cite{Jones:2002cv,Burgess:2001fx,GarciaBellido:2001ky,Shandera:2003gx} 
have highly non-trivial inflaton trajectories as a result of the brane 
interaction inside a compact manifold. 
This leaves open the possibility that more complicated models, 
like multi-throat models \cite{Iizuka:2004ct} will feature non-trivial
inflaton trajectories and therefore observable features in the power
spectrum and non-Gaussianity.

As we have shown the models have no particular problem in reproducing
the observed tilt in the power spectrum but as expected in models that
probe short trajectories, the running is small. If the observed
running is confirmed to be at the current best-fit value of $dn_s/d\ln
k \approx -0.05$ with increasing significance in future measurements
it may turn out to be the strongest constraint on these models. As
with most multi-field inflation models a detection of a gravitational
wave background would also be in conflict with the models since the
tensor perturbation are not boosted by the number of fields
contributing to inflating the universe.

\section{Acknowledgements}

The research of H.S. was supported in part by the EU under MRTN
contract MRTN-CT-2004-005104 and by PPARC under rolling grant PP/D0744X/1.
H.S. would also like to thank Oisin Mac Conamhna for discussions.  
\appendix

\section{Details of the calculation}

\subsection{The F-term potential}
\label{F_term_details} 
The F-term is calculated with the usual formula:
\begin{equation}
V_F=e^{\mathcal K}\left(K^{\overline{a}b}\overline{D_{a}W}D_{b}W-3\left|W\right|^2\right),
\end{equation}
where 
\begin{equation}
{\mathcal K} = -3\log\left(r\right).
\end{equation}
The covariant derivatives are expressed as:
\baray
D_{T}W &=& \partial_TW+W\partial_T{\mathcal K} = \partial_TW-\frac{3}{r}W\\
D_{\phi_i}W &=& \partial_{\phi_i}W+W\partial_{\phi_i}{\mathcal K} = 
\frac{\partial f_i}{\partial\phi_i}\frac{3}{r}W\\
\overline{D_{T}W} &=& \partial_{\overline{T}}\overline{W}+
\overline{W}\partial_{\overline{T}}{\mathcal K} = 
\partial_{\overline{T}}\overline{W}-\frac{3}{r}\overline{W}\\
\overline{D_{\phi_i}W} &=& \partial_{\overline{\phi_i}}\overline{W}+
\overline{W}\partial_{\overline{\phi_i}}{\mathcal K} = 
\frac{\partial f_i}{\partial\phi_i^*}\frac{3}{r}\overline{W}
\earay
The inverse K\"ahler metric needed in the expression above, 
can be calculated analytically in the general case:
\begin{equation}
{\mathcal K}^{\overline{a}b} = 
\frac{r}{3}\left(\begin{array}{ccccc}
r+\frac{\partial f_1}{\partial\varphi_1}\frac{\partial f_1}{\partial\varphi_1^*}/
\frac{\partial^2 f_1}{\partial\varphi_1\partial\varphi_1^*} + \dots +
\frac{\partial f_n}{\partial\varphi_n}\frac{\partial f_n}{\partial\varphi_n^*}/
\frac{\partial^2 f_n}{\partial\varphi_n\partial\varphi_n^*} & 
\frac{\partial f_1}{\partial\varphi_1^*}/\frac{\partial^2 f_1}{\partial\varphi_1\partial\varphi_1^*}& 
\frac{\partial f_2}{\partial\varphi_2^*}/\frac{\partial^2 f_2}{\partial\varphi_2\partial\varphi_2^*}& 
\hdots & 
\frac{\partial f_n}{\partial\varphi_n^*}/\frac{\partial^2 f_n}{\partial\varphi_n\partial\varphi_n^*}\\
\frac{\partial f_1}{\partial\varphi_1}/\frac{\partial^2 f_1}{\partial\varphi_1\partial\varphi_1^*}& 
1/\frac{\partial^2 f_1}{\partial\varphi_1\partial\varphi_1^*} & 0& \hdots & 0\\
\frac{\partial f_2}{\partial\varphi_2}/\frac{\partial^2 f_2}{\partial\varphi_2\partial\varphi_2^*} & 
0 &
1/\frac{\partial^2 f_2}{\partial\varphi_2\partial\varphi_2^*} & \hdots & 0\\
\vdots & \vdots &\vdots &\ddots &\vdots \\ 
\frac{\partial f_n}{\partial\varphi_n}/\frac{\partial^2 f_n}{\partial\varphi_n\partial\varphi_n^*} & 
0& 0& \hdots & 1/\frac{\partial^2 f_n}{\partial\varphi_n\partial\varphi_n^*}
\end{array}\right)\nonumber
\end{equation}
Using the above expression for the inverse K\"ahler metric one obtains the following terms:
\baray
K^{\overline{T}T}\overline{D_{T}W}D_{T}W &=& \frac{r}{3}
\left(\partial_{\overline{T}}\overline{W}-\frac{3}{r}\overline{W}\right)
\left(\partial_TW-\frac{3}{r}W\right)
\left(r+\frac{\frac{\partial f_1}{\partial\phi_1}\frac{\partial f_1}{\partial\phi_1^*}}
{\frac{\partial^2 f_1}{\partial\phi_1\partial\phi_1^*}} + \dots +
\frac{\frac{\partial f_n}{\partial\phi_n}\frac{\partial f_n}{\partial\phi_n^*}}
{\frac{\partial^2 f_n}{\partial\phi_n\partial\phi_n^*}}\right) \\
K^{\overline{T}\phi_i}\overline{D_{T}W}D_{\phi_i}W &=& \frac{r}{3}
\left(\partial_{\overline{T}}\overline{W}-\frac{3}{r}\overline{W}\right)W
\left(\frac{3}{r}\frac{\partial f_1}{\partial\phi_1} \frac{\frac{\partial f_1}{\partial\phi_1^*}}
{\frac{\partial^2 f_1}{\partial\phi_1\partial\phi_1^*}}
+ \dots +
\frac{3}{r}\frac{\partial f_n}{\partial\phi_n}\frac{\frac{\partial f_n}{\partial\phi_n^*}}
{\frac{\partial^2 f_n}{\partial\phi_n\partial\phi_n^*}}\right)\\
K^{\overline{\phi_i}T}\overline{D_{\phi_i}W}D_{T}W &=& \frac{r}{3}
\left(\partial_TW-\frac{3}{r}W\right)\overline{W}
\left(\frac{3}{r}\frac{\partial f_1}{\partial\phi_1^*} \frac{\frac{\partial f_1}{\partial\phi_1}}
{\frac{\partial^2 f_1}{\partial\phi_1\partial\phi_1^*}}
+ \dots +
\frac{3}{r}\frac{\partial f_n}{\partial\phi_n^*}\frac{\frac{\partial f_n}{\partial\phi_n}}
{\frac{\partial^2 f_n}{\partial\phi_n\partial\phi_n^*}}\right)\\
K^{\overline{\phi_i}\phi_j}\overline{D_{\phi_i}W}D_{\phi_j}W &=& \frac{r}{3}W\overline{W}
\left(\frac{9}{r^2}\frac{\partial f_1}{\partial\phi_1}\frac{\partial f_1}{\partial\phi_1^*}
\frac{1}
{\frac{\partial^2 f_1}{\partial\phi_1\partial\phi_1^*}} + \dots +
\frac{9}{r^2}\frac{\partial f_n}{\partial\phi_n}\frac{\partial f_n}{\partial\phi_n^*}\frac{1}
{\frac{\partial^2 f_n}{\partial\phi_n\partial\phi_n^*}}\right)
\earay

Adding all the above terms and $-3\left|W\right|^2$ we obtain the F-term:
\begin{equation}
V_F=\frac{1}{r^3}\left[\frac{r^2}{3}\partial_{\overline{T}}\overline{W}\partial_TW
\left(1+\frac{\frac{\partial f_1}{\partial\phi_1}\frac{\partial f_1}{\partial\phi_1^*}}
{r\frac{\partial^2 f_1}{\partial\phi_1\partial\phi_1^*}} + \dots +
\frac{\frac{\partial f_n}{\partial\phi_n}\frac{\partial f_n}{\partial\phi_n^*}}
{r\frac{\partial^2 f_n}{\partial\phi_n\partial\phi_n^*}}\right)-
r\left(W\partial_{\overline{T}}\overline{W} + 
\overline{W}\partial_TW\right)\right].
\end{equation}
We assumed that the superpotential itself is independent of the fields 
$\phi_1 \dots \phi_n$.

For a racetrack superpotential (\ref{racetrack_W}) the corresponding 
F-term potential becomes:
\baray
&& V_F = \frac{e^{-a\sigma}}
{3r^2\frac{\partial^2f}{\partial\phi\partial\overline{\phi}}
\frac{\partial^2g}{\partial\psi\partial\overline{\psi}}}
\left[aA^2e^{-a\sigma}\left(a\dxdyg\dxf\dyf + a\dxdyf\dxg\dyg + 
\dxdyf\dxdyg\left(6+ar\right)\right)+\right. \nonumber\\
&& \left.3aAw_0\dxdyf\dxdyg\cos\left(a\omega\right)\right] + 
\nonumber\\
&& \frac{e^{-b\sigma}}{3r^2\dxdyf\dxdyg}\left[
bB^2e^{-b\sigma}\left(b\dxdyg\dxf\dyf+b\dxdyf\dxg\dyg+
\dxdyf\dxdyg\left(6+br\right)\right)+
\right.
\nonumber\\
&&  \left.3bBw_0\dxdyf\dxdyg\cos\left(b\omega\right)\right] + 
\frac{e^{-\left(a+b\right)\sigma}}{3r^2\dxdyf\dxdyg}\left[
3AB\left(a+b\right)\dxdyf\dxdyg + \right. 
\nonumber\\
&& \left. abAB\left(\dxdyg\dxf\dyf+\dxdyf\dxg\dyg -r\dxdyf\dxdyg\right)
\right]\cos\left(\left(a-b\right)\omega\right)
\label{general_F_term}
\earay
The expression above becomes much simpler if the functions $f$ and $g$ are 
taken to be the simplest ones, $f=\phi\overline{\phi}$ and 
$g=\psi\overline{\psi}$.

\subsection{Perturbation of the kinetic term}
\label{detail_pert_kinetic}
This is the first term in Eq. (\ref{inhomogeneous_eq}): 
\begin{equation} 
\frac{1}{\sqrt{-g}}\partial_{\mu} 
\left[\sqrt{-g}g^{\mu\nu}\partial_{\nu}\phi^{K}\right],
\label{k_term}
\end{equation} 
and we expand this term as follows: 
\baray 
&& \frac{1}{\sqrt{-\det\left(g+\delta g\right)}}\partial_{\mu} 
\left[\sqrt{-\det\left(g+\delta g\right)} 
\left(g^{\mu\nu}+\delta g^{\mu\nu}\right) 
\partial_{\nu}\left(\phi^{K}+\delta\phi^{K}\right)\right] = 
\nonumber \\ 
&& \underbrace{\frac{1}{\sqrt{-g}}\partial_{\mu} 
\left[\sqrt{-g}g^{\mu\nu}\partial_{\nu}\phi^{K}\right]}_{\text{background}} + 
\frac{1}{\sqrt{-g}}\partial_{\mu} 
\left[\sqrt{-g}g^{\mu\nu}\partial_{\nu}\delta\phi^{K}\right]+ 
\frac{1}{\sqrt{-g}}\partial_{\mu} 
\left[\sqrt{-g}\delta g^{\mu\nu}\partial_{\nu}\phi^{K}\right]
\nonumber \\ 
&& + \frac{1}{\sqrt{-g}}\partial_{\mu} 
\left[\sqrt{-g} \frac{1}{2} Tr\left[g^{-1}\delta g\right]   
g^{\mu\nu}\partial_{\nu}\phi^{K}\right] 
- \frac{1}{\sqrt{-g}} \frac{1}{2} Tr\left[g^{-1}\delta g\right]  
\partial_{\mu}\left[\sqrt{-g}g^{\mu\nu}\partial_{\nu}\phi^{K}\right] 
\label{exp_k_term}
\earay 
In the above equation we used the expansion of a determinant: 
\baray 
&& \det\left[{g+\delta g}\right] =  
\det \left[g\left({\mathbbm 1}+g^{-1}\delta g\right)\right] =  
\det g \det\left[\left({\mathbbm 1}+g^{-1}\delta g\right)\right] =  
\det g e^{Tr\log\left({\mathbbm 1}+g^{-1}\delta g\right)} \simeq 
\nonumber \\ 
&& \det g e^{Tr\left[g^{-1}\delta g\right]} \simeq  
\det g\left({\mathbbm 1} + Tr\left[g^{-1}\delta g\right]\right) 
\earay 
and therefore: 
\begin{equation} 
\sqrt{-\det\left[{g+\delta g}\right]} =  
\sqrt{-\det g}\left(1+\frac{1}{2}Tr\left[g^{-1}\delta g\right]\right). 
\end{equation} 

We now proceed to simplify each one of the 4 terms in Eq.(\ref{exp_k_term}) 
above using the background metric Eq.(\ref{FRW_metric}) and the fact that the 
background fields and metric depend only on time. We write the Laplacian 
of a generic field $\Phi$ in terms of the momentum as
$\partial^{m}\partial_{m}\Phi = -k^{2}\Phi$ and proceed with the
details of the perturbations of each of the 4 terms 
in Eq.(\ref{exp_k_term}):
\subsubsection{1st term} 
This term is straightforward to compute, as it involves only the 
perturbation of the scalar field:
\baray 
&& \frac{1}{\sqrt{-g}}\partial_{\mu} 
\left[\sqrt{-g}g^{\mu\nu}\partial_{\nu}\delta\phi^{K}\right] =  
\frac{1}{a^3}\partial_{\mu}\left[a^3g^{\mu\nu} 
\partial_{\nu}\delta\phi^{K}\right] =  
\nonumber \\ 
&& -\frac{1}{a^3}\partial_{0}\left[a^3\dot{\delta\phi^{K}}\right] +  
\frac{1}{a^2}\partial_{m}\partial^{m}\delta\phi^{K} =  
 -\ddot{\delta\phi^{K}} - 3\frac{\dot{a}}{a}\dot{\delta\phi^{K}} 
-\frac{k^2}{a^2}\delta\phi^{K} 
\earay 
 
\subsubsection{2nd term} 
 In order to proceed we will need the form of the perturbations of the  
inverse metric. We can easily find the inverse metric imposing the  
condition that  
\begin{equation} 
\left(g_{\mu\nu} + \delta g_{\mu\nu}\right) 
\left(g^{\nu\lambda} + \delta g^{\nu\lambda}\right) = \delta_{\mu}^{\lambda}, 
\end{equation} 
to first order in the perturbations. We find the inverse metric to have the  
expression: 
\begin{equation} 
g^{\mu\nu} + \delta g^{\mu\nu} =  
\left(\begin{array}{cc} 
-\left(1-2A\left(t, {\bf x}\right)\right) &  
\frac{1}{a}\left(t\right)B_{, i} \left(t, {\bf x}\right) \\ 
\frac{1}{a}\left(t\right)B_{, i} \left(t, {\bf x}\right) &  
\frac{1}{a^2}\left(t\right)\left[\left(1+2\psi\left(t, {\bf x}\right)\right)\delta_{ij}- 
2E_{,ij}\left(t, {\bf x}\right)\right].
\end{array}\right) 
\end{equation} 
With this expression the 2nd kinetic term perturbation becomes: 
\baray 
&& \frac{1}{\sqrt{-g}}\partial_{\mu} 
\left[\sqrt{-g}\delta g^{\mu\nu}\partial_{\nu}\phi^{K}\right] =  
\frac{1}{a^3}\partial_{0}\left[a^3\delta g^{00} \dot{\phi^{K}}\right] + 
\frac{1}{a^3}\partial_{m}\left[a^3\delta g^{m0} \dot{\phi^{K}}\right] = 
\nonumber \\ 
&& \frac{1}{a^3}\partial_{0}\left[a^3 \left(2A\right) \dot{\phi^{K}}\right] + 
\partial_{m}\left[\frac{B_{,m}}{a} \dot{\phi^{K}}\right] =  
2A\ddot{\phi^{K}} + 6A\frac{\dot{a}}{a}\dot{\phi^{K}} + 
\dot{2A}\dot{\phi^{K}} +  
\frac{1}{a}\dot{\phi^{K}}\partial^{m}\partial_{m}B 
\earay 
 
\subsubsection{3rd term} 
This terms has the expression: 
\baray 
&& \frac{1}{\sqrt{-g}}\partial_{\mu} 
\left[\sqrt{-g} \frac{1}{2} Tr\left[g^{-1}\delta g\right]   
g^{\mu\nu}\partial_{\nu}\phi^{K}\right] =  
-\frac{1}{a^3}\partial_{0}\left[a^3 \frac{1}{2}  
Tr \left[g^{-1}\delta g\right] \dot{\phi^{K}}\right] =  
\nonumber \\ 
&& -\ddot{\phi^{K}} \frac{1}{2} Tr\left[g^{-1}\delta g\right]  
-\dot{\phi^{K}} \frac{1}{2} \dot{Tr\left[g^{-1}\delta g\right]} -  
\frac{3\dot{a}}{2a}\dot{\phi^{K}}Tr\left[g^{-1}\delta g\right]  
\earay 
where the trace of the metric perturbation has the form:: 
\begin{equation} 
\frac{1}{2}Tr\left[g^{-1}\delta g\right] =  
A - 3\psi + E_{,11} + E_{,22} + E_{,33}  = 
A - 3\psi - k^2E.
\end{equation} 
 
\subsubsection{4th term} 
Finally, this is the simplest of the 4 terms and it has the expression: 
\baray 
&& -\frac{1}{\sqrt{-g}} \frac{1}{2} Tr\left[g^{-1}\delta g\right]  
\partial_{\mu}\left[\sqrt{-g}g^{\mu\nu}\partial_{\nu}\phi^{K}\right] = 
-\frac{1}{2a^3}Tr\left[g^{-1}\delta g\right]  
\partial_{0}\left[a^3\underbrace{\left(-1\right)}_{g^{00}}\dot{\phi^{K}}\right] = 
\nonumber \\ 
&& \frac{1}{2} \ddot{\phi^{K}}Tr\left[g^{-1}\delta g\right] + 
\frac{3\dot{a}}{2a}\dot{\phi^{K}}Tr\left[g^{-1}\delta g\right] 
\earay 
We see that these two terms we obtained will cancel 2 of the terms from the  
previous expression. We can now move on to the potential terms. 

\subsection{Details of the Calculations: M-S variables} 
 
We present here the details of the calculation for the term: 
\begin{equation} 
-2A_QG^{KI}\frac{\partial V}{\partial\phi^{I}} +  
\left[\dot{A_Q} + \frac{1}{H}\frac{k^2}{a^2}\Psi \right]\dot{\phi^{K}}, 
\end{equation} 
appearing on the R.H.S. of the perturbation equation. We want it to 
take the appropriate form in terms of the Sasaki-Mukhanov variables.  
We will therefore use Eq. (\ref{Adot}) and (\ref{Psi}) in the expression 
above as well as the equation of motion for the background fields, Eq.  
(\ref{timeeq}). The expression above becomes: 
\baray 
&& -2A_QG^{KI}\frac{\partial V}{\partial\phi^{I}} +  
\left[\dot{A_Q} + \frac{1}{H}\frac{k^2}{a^2}\Psi \right]\dot{\phi^{K}} =  
-\frac{8\pi G}{H}\underbrace{G^{KI}\frac{\partial V} 
{\partial\phi^{I}}}_{\text{eq. mot.}} 
G_{LM}\dot\phi^{L}\delta\phi^{M}+ 
\nonumber \\ 
&& \frac{4\pi G}{H}\dot\phi^{K}\left[ 
-\frac{\dot H}{H}G_{IJ}\dot\phi^{I}\delta\phi^{J}+ 
\frac{\partial G_{IJ}}{\partial\phi^{L}}\dot{\phi^{I}}\dot\phi^{L}\delta{\phi^{J}} 
+G_{IJ}\ddot\phi^{I}\delta\phi^{J} 
+G_{IJ}\dot\phi^{I}\delta\dot\phi^{J}+\right. 
\nonumber \\ 
&&
\underbrace{\frac{4\pi G}{H}G_{LM}\dot\phi^{L}\dot\phi^{M} 
  G_{IJ}\dot\phi^{I}\delta\phi^{J}}_{\text{calculate } \dot H/H} 
-G_{IJ}\dot\phi^{I}\delta\dot\phi^{J} 
-\hf \frac{\partial G_{IJ}}{\partial\phi^{L}}\dot\phi^{I}\dot\phi^{J}\delta\phi^{L} 
-\underbrace{\frac{\partial V}{\partial \phi^{L}}}_{\text{eq. mot.}}\delta\phi^{L} 
\nonumber \\ 
&&\Biggl. - 3HG_{IJ}\dot\phi^{I}\delta\phi^{J} 
\Biggr] 
\earay 
We use the equation of motion to express $\frac{\partial V}{\partial \phi^{L}}$ 
in terms of the fields and their derivatives. We also notice that the terms  
$G_{IJ}\dot\phi^{I}\delta\dot\phi^{J}$ cancel and the terms containing the derivatives 
of the K\"ahler metric can be combined to form the Christoffel symbol with all indices  
lowered. 
\baray
&& \frac{\partial G_{IJ}}{\partial\phi^{L}} 
\dot{\phi^{I}}\dot\phi^{L}\delta{\phi^{J}} 
-\hf \frac{\partial G_{IJ}}{\partial\phi^{L}} 
\dot\phi^{I}\dot\phi^{J}\delta\phi^{L}= 
\nonumber \\
&& \hf\left[\frac{\partial G_{LJ}}{\partial\phi^{I}} + 
\frac{\partial G_{IL}}{\partial\phi^{J}}- 
\frac{\partial G_{IJ}}{\partial\phi^{L}}\right] 
\dot\phi^{I}\dot\phi^{J}\delta\phi^{L}= 
\Gamma_{LIJ}\dot\phi^{I}\dot\phi^{J}\delta\phi^{L} 
\earay 
Finally, the term $\frac{4\pi G}{H}G_{LM}\dot\phi^{L}\dot\phi^{M} 
G_{IJ}\dot\phi^{I}\delta\phi^{J}$ can be related to $\dot H/H$. 
We start out by calculating $\dot H/H$: 
\baray 
&& \frac{\dot H}{H} =  
\frac{1}{2\left(\hf G_{IJ}\dot\phi^{I}\dot\phi^{J} + V\right)} 
\left[G_{IJ}\ddot\phi^{I}\dot\phi^{J} +  
\frac{\partial G_{IJ}}{\partial\phi^{L}}\dot\phi^{I}\dot\phi^{J}\dot\phi^{L}+ 
\frac{\partial V}{\partial \phi^{L}}\dot\phi^{L}\right]
\nonumber \\ 
&&= \frac{1}{2\left(\hf G_{IJ}\dot\phi^{I}\dot\phi^{J} + V\right)} 
\left[-G_{IJ}\dot\phi^{J}\left(3H\dot\phi^{I}+ 
G^{IL}\frac{\partial V}{\partial \phi^{L}} +  
\Gamma^{I}_{LM}\dot\phi^{L}\dot\phi^{M}\right) \right.
\nonumber \\  
&& \left.+\frac{\partial G_{IJ}}{\partial\phi^{L}}\dot\phi^{I}\dot\phi^{J}\dot\phi^{L}+ 
\frac{\partial V}{\partial \phi^{L}}\dot\phi^{L} \right] 
\nonumber \\ 
&&=\frac{-3HG_{IJ}\dot\phi^{I}\dot\phi^{J}} 
{2\left(\hf G_{IJ}\dot\phi^{I}\dot\phi^{J} + V\right)} =  
-\frac{4\pi G HG_{IJ}\dot\phi^{I}\dot\phi^{J}}{H^2} =  
- \frac{4\pi G}{H}G_{IJ}\dot\phi^{I}\dot\phi^{J} 
\earay 
The above result tells us that we can substitute: 
\begin{equation} 
\frac{4\pi G}{H}G_{LM}\dot\phi^{L}\dot\phi^{M} 
G_{IJ}\dot\phi^{I}\delta\phi^{J} =  
-\frac{\dot H}{H} G_{IJ}\dot\phi^{I}\delta\phi^{J}.
\end{equation} 
The expression we want to calculate becomes: 
\baray 
&& -2A_QG^{KI}\frac{\partial V}{\partial\phi^{I}} +  
\left[\dot{A_Q} + \frac{1}{H}\frac{k^2}{a^2}\Psi \right]\dot{\phi^{K}} =  
\nonumber \\ 
&&\frac{8\pi G}{H}\left(\ddot\phi^{K} + 3H\dot\phi^{K} +    
\Gamma^{K}_{IJ}\dot\phi^{I}\dot\phi^{J}\right)G_{LM}\dot\phi^{L}\delta\phi^{M} 
\nonumber \\ 
&& +\frac{4\pi G}{H}\dot\phi^{K}\left[ 
-2\frac{\dot H}{H}G_{IJ}\dot\phi^{I}\delta\phi^{J} 
+\Gamma_{LIJ}\dot\phi^{I}\dot\phi^{J}\delta\phi^{L} 
+G_{IJ}\ddot\phi^{I}\delta\phi^{J}+ 
\right. 
\nonumber \\ 
&& \left.\left(\ddot\phi^{M} + 3H\dot\phi^{M} +  
\Gamma^{M}_{IJ}\dot\phi^{I}\dot\phi^{J}\right)G_{ML}\delta\phi^{L} 
-3HG_{IJ}\dot\phi^{I}\delta\phi^{J}\right] =  
\nonumber \\ 
&& \frac{8\pi G}{H}\left[ 
-\frac{\dot H}{H}G_{LM}\dot\phi^{L}\dot\phi^{M}\delta\phi^{J}+ 
G_{LM}\dot\phi^{L}\ddot\phi^{K}\delta\phi^{M} + \right. 
\nonumber \\ 
&&G_{LM}\ddot\phi^{L}\dot\phi^{K}\delta\phi^{M} + 
3HG_{LM}\dot\phi^{L}\dot\phi^{K}\delta\phi^{M} 
\nonumber \\ 
&&\left.+\Gamma_{LIJ}\dot\phi^{I}\dot\phi^{J}\delta\phi^{L}\dot\phi^{K}+ 
\Gamma^{K}_{IJ}\dot\phi^{I}\dot\phi^{J}G_{LM}\dot\phi^{L}\delta\phi^{M} 
\right] =  
\nonumber \\ 
&&  \frac{8\pi G}{a^{3}}\left[G_{IJ}\frac{\partial}{\partial t} 
\left(\frac{a^{3}\dot\phi^{I}\delta\phi^{K}}{H}\right)\right]\delta\phi^{J}+ 
\frac{8\pi G}{H}\left[\Gamma_{LIJ}\dot\phi^{I}\dot\phi^{J}\delta\phi^{L}\dot\phi^{K}+ 
\Gamma^{K}_{IJ}\dot\phi^{I}\dot\phi^{J}G_{LM}\dot\phi^{L}\delta\phi^{M} 
\right] 
\earay 
In the end we obtain the equation: 
\baray 
&&\ddot{Q^{K}} + 3H\dot{Q^{K}} +\frac{k^2}{a^2}Q^{K}  
+2\Gamma_{IJ}^{K}\dot{\phi^I}\dot{Q^J}  
+\left[G^{KI}\frac{\partial^2 V}{\partial\phi^{I}\partial\phi^{J}} +  
\frac{\partial G^{KI}}{\partial\phi^{J}}\frac{\partial V}{\partial\phi^{I}}  
\right. 
\nonumber\\  
&&+ \frac{\partial \Gamma_{IJ}^{K}}{\partial \phi^L}\dot{\phi^I}\dot{\phi^J}\dot\phi^{L} 
-\left(\frac{8 \pi G}{a^3}\right)  
G_{IJ}\frac{\partial}{\partial t} \left(\frac{a^3\dot\phi^K\dot\phi^I}{H}\right) 
\nonumber\\  
&& \left.
-\left(\frac{8 \pi G}{H}\right)\Gamma_{ILJ}\dot\phi^K\dot\phi^I\dot\phi^L 
-\left(\frac{8 \pi G}{H}\right)\Gamma^{K}_{LM}\dot\phi^{L}\dot\phi^{M}G_{IJ}\dot\phi^{I} 
\right]Q^{J} = 0 
\earay

\end{document}